\documentclass[a4paper,11pt]{article}
\usepackage{jheppub,braket} % for details on the use of the package, please see the JINST-author-manual

\title{Fluctuations in the Entropy of Hawking Radiation}

\author{Raphael Bousso$^{1,2}$ and Masamichi Miyaji$^{3,4}$}

\affiliation{$^1$ Department of Physics, 
University of California, Berkeley, CA 94720, USA}
\affiliation{$^2$ Lawrence Berkeley National Laboratory, Berkeley, CA 94720, USA}
\affiliation{$^3$ Institute for Advanced Research, Nagoya University, Nagoya, Aichi 464-8601, Japan}
\affiliation{$^4$Department of Physics, Nagoya University, Nagoya, Aichi 464-8602, Japan}
\emailAdd{bousso@berkeley.edu}
\emailAdd{masamichi.miyaji@gmail.com}

\abstract{We use the gravitational path integral (GPI) to compute the fluctuations of the Hawking radiation entropy around the Page curve, in a two-dimensional model introduced by Penington \emph{et al}. Before the Page time, we find that $\delta S = e^{-S}/\sqrt{2}$, where $S$ is the black hole entropy. This result agrees with the Haar-averaged entropy fluctuations in a bipartite system. After the Page time, we find that $\delta S \sim e^{-S}$, up to a prefactor that depends logarithmically on the width of the microcanonical energy window. This is not symmetric under exchange of subsystem sizes and so does not agree with the Haar average for a subsystem of fixed Hilbert space dimension. The discrepancy can be attributed to the fact that the black hole Hilbert space dimension is not fixed by the state preparation: even in a microcanonical ensemble with a top-hat smearing function, the GPI yields an additive fluctuation in the number of black hole states. This result, and the fact that the Page curve computed by the GPI is smooth, all point towards an ensemble interpretation of the GPI.  
}

\newcommand{\imineq}[2]{\vcenter{\hbox{\includegraphics[height=#2ex]{#1}}}}

\numberwithin{equation}{section}

\makeatletter
\gdef\@fpheader{\mbox{}}
\makeatother

\begin{document}
\maketitle
\flushbottom

\section{Introduction}

The black hole information paradox can be phrased in terms of the entropy of the Hawking radiation of an evaporating black hole. General relativity predicts a structureless horizon at late times; this implies~\cite{Hawking:1975vcx} that the radiation entropy should increase monotonically. Quantum mechanical unitarity, on the other hand, implies the Page curve~\cite{Page:1993df}. That is, the radiation entropy should increase only until it equals the black hole's Bekenstein-Hawking entropy, and thereafter it should be given by the latter.

Recently, the Page curve was derived~\cite{Penington:EntanglementWedge, Almheiri:EntropyofBulkFields} from the QES formula~\cite{Ryu:2006bv,Hubeny:2007xt,Faulkner:2013ana,Engelhardt:2014gca}, providing substantial evidence that black holes return information (and hence possess structure at the horizon~\cite{Almheiri:2012rt}). The QES formula can be viewed as a direct computation of the radiation entropy from the gravitational path integral~\cite{Lewkowycz:2013nqa}, by analytic continuation of the Renyi entropies~\cite{Penington:ReplicaWormholeWestCoast, Almheiri:ReplicaWormholeEastCoast}. 

The gravitational derivation of the Page curve has put a spotlight on a curious feature of the gravitational path integral: it appears to compute an ensemble average. This feature had already been seen explicitly in the case of Jackiw-Teitelboim (JT) gravity~\cite{Teitelboim:1983ux,Jackiw:1984je}, where Euclidean wormholes destroy factorization of the partition function of two copies of the boundary theory. Indeed, JT gravity shares low energy properties~\cite{Sarosi:2017ykf} with the Sachdev-Ye-Kitaev (SYK) model~\cite{KitaevSYK,Sachdev:1992fk}, an ensemble of quantum mechanical theories with a statistical distribution of coupling constants. 

The QES formula successfully computes the Page curve not just in JT, but in Einstein gravity in any dimension, as follows. First, the bulk geometry and state is computed semiclassically, following Hawking. Of course, this picture would predict a thermal state for the radiation, with monotonically increasing entropy. But the state is not extracted from the calculation. Instead the entropy is obtained by applying the QES prescription to the semiclassical geometry, yielding the Page curve. 

Thus, the derivation of the Page curve uses an intermediate step that is apparently inconsistent with the final answer~\cite{Bousso:2019ykv}. It is vital that the semiclassical geometry is used in the intermediate step, since the excessive entropy obtained by Hawking is precisely what causes the dominance of a nontrivial QES after the Page time.

This apparent tension is resolved if we assume that the gravitational path integral is dual to a suitable ensemble~\cite{Bousso:2020kmy}. Writing $\mathbb{E}[\ldots]$ for the ensemble average, it becomes possible that $\mathbb{E}[S(\rho)] \neq S(\mathbb{E}[\rho])$, where $\rho$ describes the radiation state produced in each unitary member of the ensemble and we have suppressed an index labelling these members. Another way of saying this is that the entropy is approximately the same in most states of the ensemble, whereas the state of the Hawking radiation depends sensitively on the theory and is not self-averaging. (Appropriate ensembles dual to gravity theories are not generally known; indeed, it is not clear whether the ensemble should consist of unitary theories with different couplings as in SYK, or of some other form of averaging.)

%\subsection{Fluctuations of the Page curve}

Here we will study a refinement of the Page curve: we will use the gravitational path integral to compute the fluctuations of the entropy of the Hawking radiation. Their magnitude was not known and is of interest in its own right. Moreover, the fact that we find a nonzero result not seen in the Page curve calculation constitutes further evidence for the ensemble interpretation of the gravitational path integral.

In a black hole evaporation process described by some specific unitary evolution, the Page curve should not be completely smooth. One expects it to exhibit small fluctuations. To see this, consider a system of $n$ qubits. In a Haar-typical quantum state, the first $k$ bits will have an entropy approximately given by the Page curve. But there are special quantum states, such as product states, which deviate drastically from the Page curve. A consistent interpolation between these facts requires that typical states deviate from the Page curve by appropriately small fluctuations. Like the Page curve itself, this argument extends to a unitarily evaporating black hole.

Our computation is done in a version of JT gravity coupled to matter, known as the PSSY model or West Coast model~\cite{Penington:ReplicaWormholeWestCoast}. We consider a black hole in the microcanonical ensemble at energy $E$, with Hilbert space dimension $e^{S(E)}$. The dimension of the radiation Hilbert space is denoted $k$. Both $E$ and $k$ can be chosen freely but are held fixed in the path integral calculation. For example, choosing $k=e^{S(E)}$ corresponds to computing the value of some quantity at the Page time. The Page curve for the entropy of the Hawking radiation $\mathbf{R}$ is given by $S_{\mathbf{R}}= \min\Big[\log k-\frac{k}{2e^{S(E)}},~S(E)-\frac{e^{S(E)}}{2k}\Big]$. 

Throughout this paper, we will use the notation
\begin{equation}
    \delta A:=\sqrt{\mathbb{E}[A^2]-\mathbb{E}[A]^2}
\end{equation}
for the fluctuation (in the ensemble implicitly dual to the gravitational path integral) of any quantity $A$.

We find that the entropy fluctuation in the large $k,~e^{S_0}$ approximation is given by
\begin{eqnarray}\label{eq:mainresult}
    \delta S_{\mathbf{R}}
    =e^{-S(E)}\times\left\{
    \begin{aligned}
    &\sqrt{\frac{1}{2}-\frac{k}{4e^{S(E)}}+\frac{\log (e^{\frac{3}{2}}\frac{\Delta E}{a})}{4\pi^2}\frac{k^2}{e^{2S(E)}}}+O(e^{2S(E)}k^{-3})&  &(1\ll e^{S(E)}-k)\\
    &\sqrt{\frac{e^{2S(E)}}{2k^2}-\frac{e^{3S(E)}}{4k^3}+\frac{\log (e^{\frac{3}{2}}\frac{\Delta E}{a})}{\pi^2}\left(1-\frac{e^{S(E)}}{2k}\right)^2}+O(e^{-S(E)})&  &(k-e^{S(E)}\gg1).
    \end{aligned}
    \right.
\end{eqnarray}
Here $\Delta E$ is the width of the microcanonical energy window, whose edges are smeared by $a\ll \Delta E$ so that the energy window becomes continuous. $\Delta E$ is chosen to be an $O(1)$ quantity in the $e^{-S_0}$ expansion, which allows us to neglect higher order terms in the genus expansion.

A puzzling aspect of this result is that it is not symmetric under interchange of $e^{S(E)}$ and $k$. By contrast, the Page curve (including its fluctuations) are manifestly symmetric, since it does not matter which subsystem dimension we label $e^S$ and which we label $k$. For example, given that $\delta S_{\mathbf{R}}\sim e^{-S(E)}$ for $k\ll e^{S(E)}$, symmetry would require $\delta S_{\mathbf{R}}\sim 1/k$ for $k\gg e^{S(E)}$, which is much smaller than the result we obtain, $e^{-S(E)}$.

Again, this apparent contradiction is resolved by assuming that the gravitational path integral averages over an ensemble of theories. The black hole Hilbert space, of dimension $e^{S(E)}$, is defined by the number of black hole states in a given energy band. The precise value of this integer depends on the detailed spectrum, so it will not be the same in every theory in the ensemble. 
The black hole Hilbert space dimension controls the $k\gg S(E)$ regime of the Page curve. Our result above implies that \emph{its} fluctuation dominates over the intrinsic fluctuations in the Page curve in this regime. Specifically, the lower part of Eq.~\eqref{eq:mainresult} indicates that $\delta[e^{S(E)}]=\sqrt{\log (e^{\frac{3}{2}}\frac{\Delta E}{a})}/\pi$ to leading order, which depends logarithmically on the width $\Delta E$ of the microcanonical window. 

To verify that this is what happens, we will also compute this fluctuation directly. Because the full state is pure, $\delta[e^{S(E)}]=\delta\text{Tr}[\hat{\rho}_{\mathbf{R}}^{0+}]$. The latter quantity, the rank of the radiation density operator, can again be computed using the gravitational path integral. We find 
\begin{equation}
    \delta\text{Tr}[\hat{\rho}_{\mathbf{R}}^{0+}]
    =\left\{
    \begin{aligned}
    &~\,0 ~~ +O(k^{-1})&  &(1\ll e^{S(E)}-k)\\
    & \frac{1}{\pi}\sqrt{\log (e^{\frac{3}{2}}\frac{\Delta E}{a})}+O(e^{-S(E)})&  &(k-e^{S(E)}\gg1)~.
    \end{aligned}
    \right.
\end{equation}

%The regime $k\ll e^{S(E)}$ of our result may also seem surprising, since it is much smaller than previously established upper bounds~\cite{Patrick2006} on the entropy fluctuations of a subsystems of a system in a random state. This shows that the known upper bound was not tight. Indeed, our gravitational path integral result can be  confirmed by explicitly computing the fluctuation of the subsystem entropy in quantum mechanics, for large Hilbert space dimensions, using random matrix techniques.\footnote{This result was found by Douglas Stanford (private communication).} 

In the course of the computation, we demonstrate diagrammatically the independence of the resolvent two-point function from details of matrix integral potential, in the microcanonical PSSY model. This independence is simple to understand from the matrix integral view point, yet remains mysterious from a diagrammatic viewpoint~\cite{Brezin}. Since our method does not refer to a specific potential, we expect that it can be used for a matrix integral with arbitrary potential.

The content of this paper is as follows. In Sec.~\ref{section:Review}, we review the PSSY model and the computation of the Page curve by analytic continuation of the Renyi entropies of the Hawking radiation. In Sec.~\ref{section:non-tubular}, we compute the fluctuations around the Page curve in the PSSY model. In particular, we identify a sector in the PSSY model where entropy fluctuations are identical to those of the random bi-partite pure state. We discuss our results in Sec.~\ref{section:random state}. 

\section{PSSY Model}\label{section:Review}

In this section, we consider a version of Jackiw-Teitelboim gravity first introduced by Penington \emph{et al.}\ (PSSY)~\cite{Penington:ReplicaWormholeWestCoast}. In Sec.~\ref{2.1}, we will review the PSSY model. In Sec.~\ref{2.2} we will define the microcanonical ensemble of width $\Delta E$ around some energy. We introduce a smooth smearing function that becomes top-hat in a limit. (Obtaining the top hat as a limit will be important in this paper, whereas in earlier work the top hat could be used directly.)  In Sec.~\ref{2.3}, we review the Page curve result, and we compute the rank of the radiation density matrix.

\subsection{Action and Canonical Ensemble}
\label{2.1}

The PSSY model consists of JT gravity \cite{Teitelboim:1983ux, Jackiw:1984je, Maldacena:2016upp, Stanford:2017thb, Yang:2018gdb, Saad:2019lba,Stanford:2019vob, Saad:2019pqd} with a matter sector given by an end-of-the-world (EOW) brane with $k$ flavors and tension $\mu~(\geq0)$, anchored at the boundary. The action is
\begin{equation}
    S=S_{\text{JT}}+S_{\text{Brane}},
\end{equation}
where
\begin{equation}
    S_{\text{JT}}=-\frac{S_0}{4\pi}\left(\int_{\mathcal{M}} \sqrt{g}R+2\int_{\partial\mathcal{M}}\sqrt{h}K\right)
    -
    \frac{1}{2}\left(\int_{\mathcal{M}} \sqrt{g}\phi(R+2)+\int_{\partial\mathcal{M}}\sqrt{h}\phi K\right),
\end{equation}
\begin{equation}
    S_{\text{brane}}=\mu\int_{\text{Brane}}ds.
\end{equation}
We impose the standard asymptotic boundary condition
\begin{equation}
    ds^2|_{\partial{\mathcal{M}}}=\frac{d\tau^2}{\epsilon^2},~\phi|_{\partial{\mathcal{M}}}=\frac{1}{\epsilon}.\end{equation}
Here $\tau$ is the boundary Euclidean time. Note that this model does not contain loops of EOW branes that are not anchored at the boundary. The model is dual to an ensemble of boundary Hamiltonians; see appendix D in \cite{Penington:ReplicaWormholeWestCoast}.

One considers a state in which the EOW brane is maximally entangled with an auxiliary nongravitating system $\mathbf{R}$:
%The unnormalized global pure state is
\begin{equation}\label{eq:state}
|\Psi\rangle=k^{-1/2}\sum_{i=1}^k|i\rangle_{\mathbf{R}}|i\rangle_{\mathbf{EOW}}~.
\end{equation}
This is a toy model for the semiclassical state of a black hole whose interior is entangled with Hawking radiation. The boundary description of the bulk state is given by  
\begin{equation}\label{eq:bulkstate}
    |i\rangle_{\bold{EOW}}
    \propto\sum_{s}\sqrt{f(E_s)}2^{1/2-\mu}\Gamma[\mu-1/2+i\sqrt{2E_s}]C_{is}|E_s\rangle,
\end{equation}
where $|E_s\rangle$ are eigenstates of single instance of Hamiltonian ensemble of the matrix integral dual to the JT gravity, and Note that the state (\ref{eq:state}) is normalized only in the ensemble-averaged sense: $\mathbb{E}[\langle\Psi|\Psi\rangle]=1$, but $\mathbb{E}[\langle\Psi|\Psi\rangle^2]\neq 1$. The fluctuation of the normalization is not relevant here, but we will need to take it into account in the next section.

The reduced density matrix of the radiation is
\begin{equation}\label{eq:reduceddensity}
    \rho_{\mathbf{R}}=k^{-1}\sum_{i,j=1}^k|i\rangle\langle j|_{\mathbf{R}}\langle j|i\rangle_{\mathbf{EOW}}~.
\end{equation}
The eigenvalue density $D(\lambda)$ of $\rho_{\mathbf{R}}$ is encoded in the resolvent $R$
\begin{equation}
    D(\lambda)=\frac{R(\lambda-i\epsilon)-R(\lambda+i\epsilon)}{2\pi i}~,
\end{equation}
where
\begin{equation}
    R_{ij}(\lambda)=\langle i|\frac{1}{\lambda-\rho_{\mathbf{R}}}|j\rangle_{\mathbf{R}}~,
\end{equation}
and
\begin{equation}
    R(\lambda):=\sum_{i}R_{ii}(\lambda)=\frac{k}{\lambda}+\sum_{n=1}^{\infty}\frac{\text{Tr}[\rho_{\mathbf{R}}^n]}{\lambda^{n+1}}=
    \text{Tr}\frac{1}{\lambda-\rho_{\mathbf{R}}}.
\end{equation}

When $k$ and $e^{S_0}$ are both large, the resolvent can be computed from the gravitational path integral by only considering planar geometries. One finds the following recursion relation:
\begin{equation}
    R_{ij}(\lambda)=\frac{1}{\lambda}\delta_{ij}+\frac{1}{\lambda}\sum_{n=1}^{\infty}\frac{Z_{\text{Disk}}^{(n)}}{(kZ_{\text{Disk}}^{(1)})^n}R(\lambda)^{n-1}R_{ij}(\lambda)~,\label{eq:recursion}
\end{equation}
where $Z_{\text{Disk}}^{(n)}$ is the single-topological-disk contribution to the JT gravity partition function for $n$ boundaries, with fixed boundary condition. (Thus the $n$ AdS boundary segments are connected by $n$ EOW branes to form a single boundary.) In the canonical ensemble, each boundary has a fixed length $x_i=\beta/2$ corresponding to the inverse temperature $\beta$; thus
% , with action
% \begin{equation}
%     S_{\text{EOW}}=\mu l,
% \end{equation}
% where $l$ is the regularized length of the EOW brane. 
%The disk topology partition function is given by
\begin{multline}
    Z^{(n)}_{\text{Disk}}\left[\text{canonical ensemble,~temperature}~ \beta^{-1}~\text{on each boundary}\right]\\=e^{S_0}\int_0^{\infty}dE\, D_{\text{Disk}}(E)\, h(E,\mu)^n\, e^{-n\beta E/2},
\end{multline}
where
\begin{equation}
    D_{\text{Disk}}(E)=\frac{\sinh(2\pi\sqrt{2E})}{2\pi^2},~
    h(E,\mu):=\frac{|\Gamma(\mu-1/2+i\sqrt{2E})|^2}{2^{2\mu-1}}.
\end{equation}

\subsection{Microcanonical Ensemble and Smearing Function}
\label{2.2}

In this paper, we will focus on the microcanonical ensemble. From the canonical partition function, one can obtain the density of states by the inverse Laplace transform. By applying a top-hat smearing function $g_{(E,\Delta E)}(\tilde{E})=\theta(\Delta E/2-|\tilde{E}-E|)$ to the density of states, one would obtain the microcanonical ensemble in the sharp energy window $[E-\Delta E/2,E+\Delta E/2]$. Here $\theta(x)=1~(x>0)$, $\theta(x)=0~(x<0)$ is the step function. However, for a top-hat smearing function, the two-point function of partition functions is sensitive to fine-grained spectrum of the theory, so it depends on higher genus contributions in the GPI. 

Thus, instead of using top-hat smearing function, we will use a smooth microcanonical smearing function $f_{(E,\Delta E,a)}(\tilde{E})$ that limits to $g_{(E,\Delta E)}(\tilde{E})$ for small $a$. As long as we keep $e^{-S_0}\ll a\ll1$, we expect that the two-point function of partition functions does not receive large contributions from higher genus contributions in the GPI.

The conditions for $f_{(E,\Delta E,a)}(\tilde{E})$ are
\begin{eqnarray}\label{eq:smearingconditions}
    \left\{
    \begin{aligned}
    &0\leq f_{(E,\Delta E,a)}(\tilde{E})\leq 1,    \\
    &\underset{a\rightarrow +0}{\lim} f_{(E,\Delta E,a)}(\tilde{E})=g_{(E,\Delta E)}(\tilde{E})    .
    \end{aligned}
    \right.
\end{eqnarray}
An explicit example we will use is the trapezoid function which is continuous
\begin{equation}\label{eq:smearingappendix}
    f_{(E,\Delta E,a)}(\tilde{E})
    =\left\{
    \begin{aligned}
    &-\frac{\tilde{E}-(E+\Delta E/2+a)}{a}& &(E+\Delta E/2<\tilde{E}<E+\Delta E/2+a)\\
    &1&  &(E-\Delta E/2<\tilde{E}<E+\Delta E/2 )\\
    &\frac{\tilde{E}-(E-\Delta E/2-a)}{a}&  &(E+\Delta E/2-a<\tilde{E}<E-\Delta E)\\
    &0&  &(\text{otherwise}).
    \end{aligned}
    \right.
\end{equation}
The corresponding microcanonical disk $n$-boundary partition function is
\begin{eqnarray}
    &&Z^{(n)}_{\text{Disk}}\left[\text{microcanonical}, \text{energy}=E, \text{width}=\Delta E\right]
    \nonumber\\
    &:=&
    \Pi_{i=1}^{n}
    \left[
    \int_0^{\infty}dE_i f_{(E,\Delta E,a)}(E_i)
    \int_{x_i\in \gamma+i\mathbb{R}}dx_i
    e^{x_iE_i}\right]
    Z^{(n)}_{\text{Disk}}[\text{canonical},\text{boundary length}=x_i]
    \nonumber\\&=&
    e^{S_0}\int_0^{\infty}d\tilde{E}D_{\text{Disk}}(\tilde{E})
    \left[f_{(E,\Delta E,a)}(\tilde{E})\right]^n
    h(\tilde{E},\mu)^{n}.
\end{eqnarray}
In the limit $\Delta  E\ll 1$ and small $a$, as we explain in detail in appendix \ref{JTdisk}, we have
\begin{equation}
    \label{eq:disk}
    Z^{(n)}_{\text{Disk}}\left[\text{microcanonical}, \text{energy}=E, \text{width}=\Delta E\right]
    =e^{S(E)}h(E,\mu)^{n}+O(e^{S_0}(\Delta E)^3).
\end{equation}
Here we defined
\begin{equation}
    e^{S(E)}:=e^{S_0}D_{\text{Disk}}(E)\Delta E,
\end{equation}
which is the number of states in the microcanonical window with width $\Delta E$. We assume that $1\ll e^{S_0}\Delta E\ll e^{S_0}$, so that the second term is subleading. Note that the result (\ref{eq:disk}) does not depend on the details of the smooth microcanonical smearing function, so long as the function satisfies (\ref{eq:smearingconditions}) and $a$ being small.\footnote{The smearing function $f(\tilde{E})=\exp(-(E-\tilde{E})^2/(4\Delta E)^2)$ used in \cite{Saad:2018bqo} would not give (\ref{eq:disk}), because it violates the second condition of (\ref{eq:smearingconditions}); that is, it does not limit to the sharp microcanonical smearing function $g_{(E,\Delta E)}(\tilde{E})$.}

    \subsection{Entropy and Rank of the Radiation State}
\label{2.3}

In the following, we will assume that $a$ is small. The calculation below does not depend on whether our smearing function is sharp or smooth with $a\rightarrow 0$, unlike the entropy fluctuation considered in the next section. The sharp smearing function was used in \cite{Penington:ReplicaWormholeWestCoast}. With Eq.~\eqref{eq:disk}, the trace of Eq.~(\ref{eq:recursion}) yields a quadratic equation independent of $\mu$ for $R(\lambda)$
\begin{equation}
    R(\lambda)^2+R(\lambda)\left(\frac{e^{S(E)}-k}{\lambda}-ke^{S(E)}\right)+\frac{k^2e^{S(E)}}{\lambda}=0.
\end{equation}
The limiting behavior $R(\lambda)\underset{|\lambda|\gg 1}{\rightarrow}k/\lambda$ dictates that
\begin{eqnarray}\label{eq:resolvent}
    R(\lambda)=ke^{S(E)}\Big(\frac{1}{2}-\frac{w}{2x}-\frac{\sqrt{\frac{w^2}{4}+\frac{x^2}{4}-x(1+\frac{w}{2})}}{x}\Big),
\end{eqnarray}
where we defined $w:=\frac{e^{S(E)}-k}{k}$ and $x:=e^{S(E)}\lambda$. The density of eigenvalues is 
\begin{eqnarray}\label{eq:density}
    D(\lambda)=\frac{ke^{S(E)}}{2\pi \lambda}\sqrt{(\lambda-\lambda_-)(\lambda_+-\lambda)}+(k-e^{S(E)})\delta(\lambda)\theta(k-e^{S(E)}),
\end{eqnarray}
where $\lambda_{\pm}=e^{-S(E)}(2+w\pm 2\sqrt{1+w})$ give the endpoints of the continuous spectrum. We have $\int d\lambda D(\lambda)\lambda=1$ for normalization of $\rho_{\mathbf{R}}$ and $\int du D(u)=k$ for the rank of $\rho_{\mathbf{R}}$. This entanglement spectrum is identical to that of the reduced density matrix on $\mathbb{C}^k$ of random state on $\mathbb{C}^{k}\otimes\mathbb{C}^{e^{S(E)}}$. Thus the Renyi entropy is identical to that of the random state. The Renyi entropy $(n>0,n\neq 1)$ is \cite{Kawabata:2021hac}
\begin{eqnarray}
    S^{(n)}_{\mathbf{R}}
    &=&
    \frac{1}{1-n}\log\int d\lambda D(\lambda)\lambda^n
    \nonumber\\
    &=&
    \log\frac{k}{(1-\sqrt{\frac{k}{e^{S(E)}}})^2}
    +\frac{1}{1-n}
    \log
    {}_2F_1\left(1-n,\frac{3}{2},3,-\frac{4\sqrt{\frac{k}{e^{S(E)}}}}
    {(1-\sqrt{\frac{k}{e^{S(E)}}})^2}\right).
\end{eqnarray}
In particular, one obtains the Page curve for the von Neumann entropy of the radiation:
\begin{equation}\label{eq:Pagecurve}
    S_{\mathbf{R}}=\left\{
    \begin{aligned}
    &\log k-\frac{k}{2e^{S(E)}}+O(k^{-1}) &  (k&<e^{S(E)})\\
    &S(E)-\frac{e^{S(E)}}{2k}+O(e^{-S(E)}) &  (k&>e^{S(E)}).
    \end{aligned}
    \right.
\end{equation}
Note that (\ref{eq:Pagecurve}) is identical to the entanglement entropy of a random pure state for large Hilbert space dimensions \cite{Page:1993df}\cite{Foong:, Sen:1996ph}. The rank of the reduced density matrix is
\begin{equation}\label{eq:Pagecurveofrank}
    \text{Tr}[\rho_{\mathbf{R}}^{0+}]=\left\{
    \begin{aligned}
    &k+O(1) &  (k&<e^{S(E)})\\
    &e^{S(E)}+O(1) &  (k&>e^{S(E)}).
    \end{aligned}
    \right.
\end{equation}
%%%%%%%%%%%%%%%%%%%%%%%%%%%%%%%%%%%%%%%%%%%%%%%%%%%%%%%%%%%%%%%%%%%%%%%%%%%%%%%%%%%%%%%%%%%%%%%%%%%%%%%%%%%%%%%%%%%%%%%%%%%%%%%%%%%%%%%%%%%%%%%%%%%%%%%%%%%%%%%%%%%%%%%%%%%%%%%%%%%%%%%%%%%%%%%%%%%%%%%%%%%%%%%%%%%%%%%%%%%%%%%%%%%%%%%%%%%%%%%%%%%%%%%%%%%%%%%%%%%%%%%%%%%%%%%%%%%%%%%%%%%%%%%%%%%%%%%%%%%%%%%%%%%%%%%%%%%%%%%%%%%%%%%%%%%%%%%%%%%%%%%%%%%%%%%%%%%%%%%%%%%%%%%%%%%%%%%%%%%%%%%%%%%%%%%%%%%%%%%%%%%%%%%%%%%%%%%%%%%%%%%%%%%%%%%%%%%%%%%%%%%%%%%%%%%%%%%%%%%%%%%%

\section{Entropy Fluctuation in the PSSY Model}\label{section:non-tubular}

In this section, we compute the fluctuation of the entropy of the Hawking radiation, $\delta S_{\mathbf{R}}$, in the the PSSY model. We begin by setting up the calculation in Sec.~\ref{compute}. The leading diagrammatic contributions come from two types of topologies. We consider them separately in Sections \ref{subsection:nontubular} 
and \ref{section:tubular}, and we combine their contributions in Sec.~\ref{sum}.

\subsection{Setting Up the Calculation}
\label{compute}

At this order, it is important to note that the normalization of $\rho_{\mathbf{R}}$ defined in Eq.~\eqref{eq:reduceddensity} fluctuates. To avoid artifacts in $\delta S_{\mathbf{R}}$, we must take care to compute $S_{\mathbf{R}}$ from the normalized reduced density matrix
\begin{equation}
    \hat{\rho}_{\mathbf{R}}:=\frac{\rho_{\mathbf{R}}}{\text{Tr}[\rho_{\mathbf{R}}]}~.
\end{equation}
The entanglement spectrum of $\hat{\rho}_{\mathbf{R}}$, written as a density of eigenvalues, will be denoted $\hat{D}(\lambda)$. The associated resolvent $\hat{R}(\lambda):=\text{Tr}\frac{1}{\lambda-\hat{\rho}_{\mathbf{R}}}$ is equal to $R(\lambda)$ at leading order in $e^{-S_0}$, so the distinction between $\hat R(\lambda)$ and $R(\lambda)$ will not be important for us. However, below we introduce the resolvent two-point function, to which the normalization fluctuation would contribute at the leading order.

\begin{figure}[t]
 \begin{center}
 \includegraphics[width=11cm,clip]{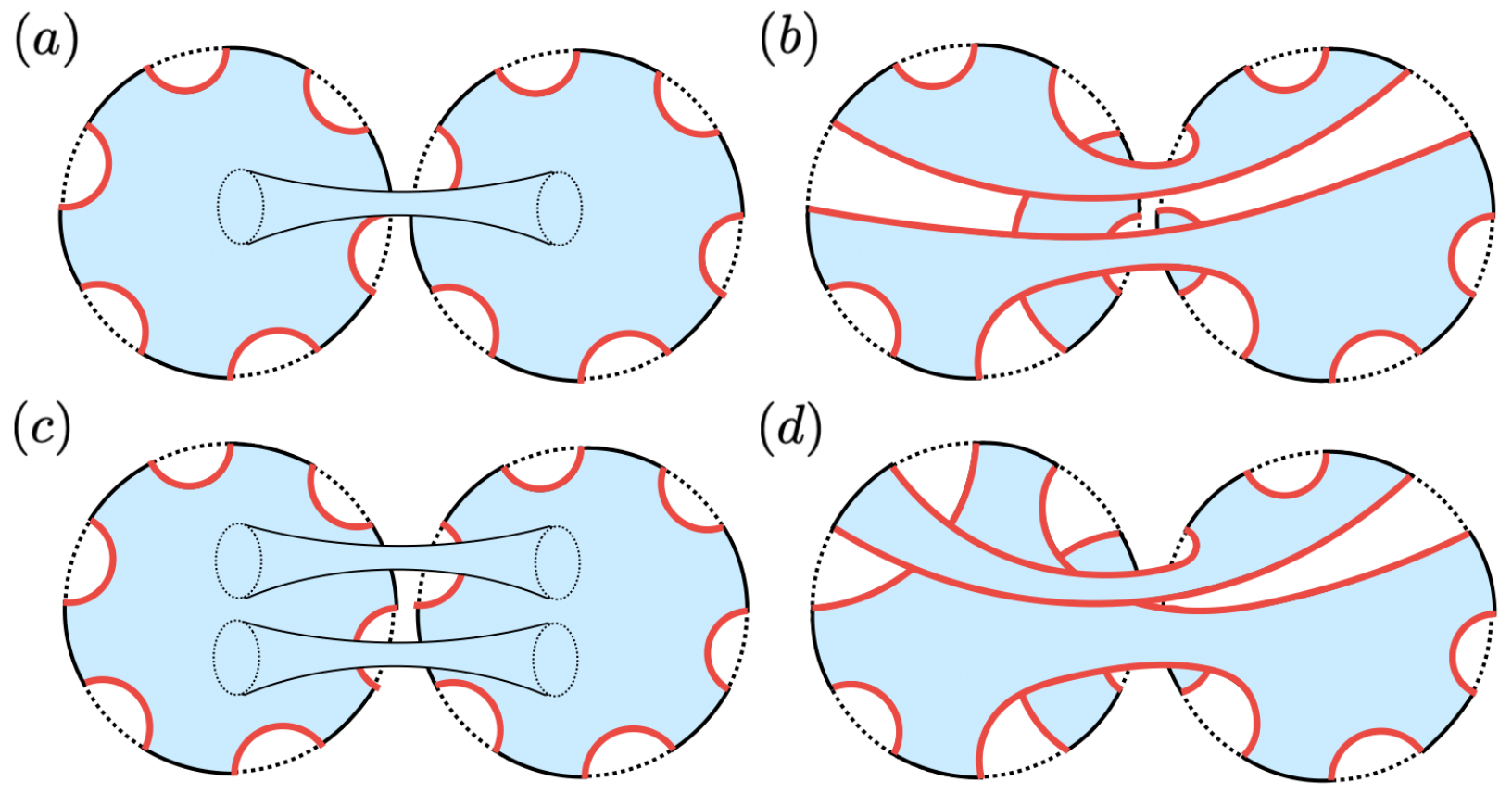}
 \end{center}
 \caption{Examples of geometries that contribute to $\hat{R}(\lambda_1,\lambda_2)$. \emph{(a)} and \emph{(b)} contribute at leading order, while \emph{(c)} and \emph{(d)} are subleading at large $e^{S(E)}$ and large $k$. \emph{(a), (c)} contain double trumpets; \emph{(c)} contains a handle; \emph{(d)} is non-planar; and \emph{(b)} and \emph{(d)} contain neither double trumpet nor handle.}
 \label{fig:wormholes}
 \end{figure}

The fluctuation of the entropy can be obtained via 
\begin{equation}
(\delta S_{\mathbf{R}})^2=\partial_{n_1}\partial_{n_2}\hat{R}_{n_1n_2}|_{n_1=n_2=1}~,
\end{equation} 
where
\begin{equation}\label{eq:defRnm}
    \hat{R}_{n_1n_2}:=\mathbb{E}[
    \text{Tr}[\hat{\rho}_{\mathbf{R}}^{n_1}]
    \text{Tr}[\hat{\rho}_{\mathbf{R}}^{n_2}]]
    -
    \mathbb{E}[\text{Tr}[\hat{\rho}_{\mathbf{R}}^{n_1}]]
    \mathbb{E}[\text{Tr}\hat{\rho}_{\mathbf{R}}^{n_2}]].
\end{equation}
It is convenient to consider the connected part of the resolvent two-point function
\begin{equation}\label{eq:connectedtwopoint}
    \hat{R}(\lambda_1,\lambda_2):=\sum_{n_1,n_2=0}^{\infty}
    \frac{\hat{R}_{n_1n_2}}{\lambda_1^{n_1+1}\lambda_2^{n_2+1}}=\mathbb{E}[
    \text{Tr}\frac{1}{\lambda_1-\hat{\rho}_{\mathbf{R}}}
    \text{Tr}\frac{1}{\lambda_2-\hat{\rho}_{\mathbf{R}}}]
    -\mathbb{E}[
    \text{Tr}\frac{1}{\lambda_1-\hat{\rho}_{\mathbf{R}}}]
    \mathbb{E}[
    \text{Tr}\frac{1}{\lambda_2-\hat{\rho}_{\mathbf{R}}}],
\end{equation}
whose discontinuity across its cut encodes the two-point function $\hat{D}(\lambda_1,\lambda_2)$ of the entanglement spectrum $\hat{D}(\lambda)$ of $\hat{\rho}_{\mathbf{R}}$ via
\begin{eqnarray}
    &&\hat{D}(\lambda_1,\lambda_2)
    \nonumber\\&&=
    -\frac{\hat{R}(\lambda_1-i\epsilon,\lambda_2-i\epsilon)-\hat{R}(\lambda_1-i\epsilon,\lambda_2+i\epsilon)-\hat{R}(\lambda_1+i\epsilon,\lambda_2-i\epsilon)+\hat{R}(\lambda_1+i\epsilon,\lambda_2+i\epsilon)}{4\pi^2 }.\nonumber\\
\end{eqnarray}
In the following, we will evaluate $\hat{R}(\lambda_1,\lambda_2)$ in the PSSY model. We can divide the geometries contributing to $\hat{R}(\lambda_1,\lambda_2)$ into two classes, by whether they include handles or an annulus topology (with possible handles) called double trumpet. Geometries containing neither a double trumpet nor a handle will be called \emph{non-tubular wormholes}. Those containing a double trumpet or handles will be called \emph{tubular wormholes}. 
See Fig.~\ref{fig:wormholes} for geometries that contribute to the resolvent two-point function.

We will evaluate $\hat{R}(\lambda_1,\lambda_2)$ diagrammatically. To avoid overcounting, it is convenient to consider the quantity
\begin{equation}
    U(\lambda):=-\sum_{n=1}^{\infty}\frac{\text{Tr}[\hat{\rho}_{\mathbf{R}}^n]}{n\lambda^{n+1}},
\end{equation}
which is related to the resolvent by
\begin{equation}
    \hat{R}(\lambda)=\partial_{\lambda}(\lambda U(\lambda))+\frac{k}{\lambda}.
\end{equation}
We will compute the connected two-point function of $U$,
\begin{equation}
    U(\lambda_1,\lambda_2):=\mathbb{E}[U(\lambda_1)U(\lambda_2)]-\mathbb{E}[U(\lambda_1)]\mathbb{E}[U(\lambda_2)],
\end{equation}
from which we can then easily deduce the resolvent two-point function, via
\begin{equation}
    \hat{R}(\lambda_1,\lambda_2)=\partial_{\lambda_1}\partial_{\lambda_2}(\lambda_1\lambda_2U(\lambda_1,\lambda_2)).
\end{equation}

%%%%%%%%%%%%%%%%%%%%%%%%%%%%%%%%%%%%%%%%%%%%%%%%%%%%%%%%%%%%%%%%%%%%%%%%%%%%%%%%%%%%%%%%%%%%%%%%%%%%%%%%%%%%%%%%%%%%%%%%%%%%%%%%%%%%%%%%%%

\subsection{Non-tubular Wormholes}\label{subsection:nontubular}

Non-tubular wormholes consist of re-connections of EOW branes without containing double trumpet geometries nor handles. We denote contributions from non-tubular wormholes by a superscript or subscript, such as $D(\lambda_1,\lambda_2)^{\text{Non-tubular}},~(\delta S_{\mathbf{R}})_{\text{Non-Tubular}}$.

\begin{figure}[t]
 \begin{center}
 \includegraphics[width=4.5cm,clip]{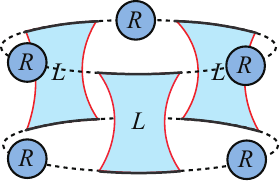}
 \end{center}
 \caption{Non-tubular wormholes connecting $\text{Tr}\rho_{\mathbf{R}}^n$ and $\text{Tr}\rho_{\mathbf{R}}^m$ in the first term of (\ref{eq:UUnontubular}). Solid red lines are EOW branes, and the black dotted lines represent flavor index contractions. $R$ represents the resolvent one point function.}
 \label{fig:non-tubular}
 \end{figure}

We first study the general situation using diagrammatic expansion. The resolvent two-point function for the random matrix with quartic potential was computed using a diagrammatic expansion in \cite{Brezin}. The expression given in this section applies to arbitrary potential. In order to sum over all possible non-tubular wormholes connecting $R(\lambda_1)$ and $R(\lambda_2)$, it is convenient to define a sum of \emph{irreducible ladder diagrams}
\begin{eqnarray}
    L^{i_1j_1}_{i_2j_2}(\lambda_1,\lambda_2)
    &:=&
    \imineq{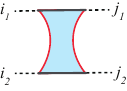}{10}
    +\imineq{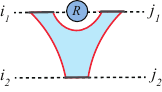}{10}
    +\cdots+
    \imineq{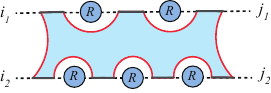}{10}+\cdots
    \nonumber\\
    &=&\delta_{i_1i_2}\delta_{j_1j_2}\sum_{n,m=1}^{\infty}\frac{Z_{\text{Disk}}^{(n+m)}R(\lambda_1)^{n-1}R(\lambda_2)^{m-1}}{(kZ_{\text{Disk}}^{(1)})^{n+m}}\nonumber\\
    &=&\delta_{i_1i_2}\delta_{j_1j_2}L(\lambda_1,\lambda_2).
\end{eqnarray}
where $L$ is defined through the last equality. Then we show in appendix \ref{appendix:proof} that
\begin{eqnarray}
\label{eq:UUnontubular}
    &&U(\lambda_1,\lambda_2)^{\text{Non-tubular}}
    =
    \frac{1}{\lambda_1\lambda_2}\sum_{n=1}^{\infty}\frac{1}{n}\Big(\sum_{i,j}R_{ij}(\lambda_1)R_{ji}(\lambda_2)\Big)^nL(\lambda_1,\lambda_2)^n\nonumber\\
    &-&
    \frac{1}{\lambda_1}\sum_{n=1}^{\infty}\frac{Z_{\text{Disk}}^{(n+1)}R(\lambda_1)^{n}}{(kZ_{\text{Disk}}^{(1)})^{n+1}}R(\lambda_2)    
    -R(\lambda_1)\frac{1}{\lambda_2}\sum_{n=1}^{\infty}\frac{Z_{\text{Disk}}^{(n+1)}R(\lambda_2)^{n}}{(kZ_{\text{Disk}}^{(1)})^{n+1}}
    +    
    R(\lambda_1)
    R(\lambda_2)
    \frac{kZ_{\text{Disk}}^{(2)}}{(kZ_{\text{Disk}}^{(1)})^2}.\nonumber\\
\end{eqnarray}
The first term in (\ref{eq:UUnontubular}) corresponds to the wormhole exchanges between $\text{Tr}\rho_{\mathbf{R}}^n$ and $\text{Tr}\rho_{\mathbf{R}}^m$, see Fig.~\ref{fig:non-tubular}.\footnote{Note that we have $R_{ij}(\lambda)=\delta_{ij} R(\lambda)/k$, which leads to a simplification $\sum_{i,j}R_{ij}(\lambda_1)R_{ji}(\lambda_2)=R(\lambda_1)R(\lambda_2)/k$.}
The second and third terms correspond to those between $\text{Tr}\rho_{\mathbf{R}}^n$ and $(\text{Tr}[\rho_{\mathbf{R}}])^m$ or $\text{Tr}\rho_{\mathbf{R}}^m$ and $(\text{Tr}[\rho_{\mathbf{R}}])^n$. The forth term corresponds to those between $(\text{Tr}[\rho_{\mathbf{R}}])^n$ and $(\text{Tr}[\rho_{\mathbf{R}}])^m$.

%%%%%%%%%%%%%%%%%%%%%%%%%%%%%%%%%%%%%%%%%%%%%%%%%%%%%%%%%%%%%%%%%%%%%%%%%%%%%%%%%%%%%%%%%%%%%%%%%%%%%%%%%%%%%%%%%%%%%%%%%%%%%%%%%%%%%%%%%%%%%%%%%%%%%%%%%%%%
%%%%%%%%%%%%%%%%%%%%%%%%%%%%%%%%%%%%%%%%%%%%
\paragraph{Microcanonical Ensemble}
We now consider the microcanonical ensemble. Using $Z_n=e^{S(E)}h^n$, one finds
\begin{equation}
    L(\lambda_1,\lambda_2)=e^{S(E)}\frac{1}{ke^{S(E)}-R(\lambda_1)}\frac{1}{ke^{S(E)}-R(\lambda_2)}.
\end{equation}
Therefore,
\begin{equation}
    U(\lambda_1,\lambda_2)^{\text{Non-tubular}}=\frac{1}{\lambda_1\lambda_2}\sum_{n=2}^{\infty}\frac{1}{n}\Big(\frac{1}{ke^{S(E)}}(\lambda_1 R(\lambda_1)-k)(\lambda_2 R(\lambda_2)-k)\Big)^n+\frac{k}{e^{S(E)}}\frac{1}{\lambda_1\lambda_2}.
\end{equation}

We now define $x:=e^{S(E)}\lambda$, $a_{\pm}:=e^{S(E)}\lambda_{\pm}$, and $\sigma(x,a_{\pm}):=(x-a_-)(x-a_+)$. One finds by explicit computation that
\begin{eqnarray}\label{eq:relationtorandomstate}
    &&\partial_{\lambda_1}\partial_{\lambda_2}(\lambda_1\lambda_2U(\lambda_1,\lambda_2)^{\text{Non-tubular}})
    \nonumber\\&&
    =e^{2S(E)}R^x(x_1,x_2:2,a_{\pm})
    -\frac{1}{ke^{S(E)}}
    \partial_{x_1}\Big(x_1R^x(x_1)\Big)
    \partial_{x_2}\Big(x_2R^x(x_2)\Big).\nonumber\\
\end{eqnarray}
where $R^x(x):=k^{-1}e^{-S(E)}R(\lambda)$ and \cite{Stanford:2019vob}
\begin{equation}\label{eq:twopoint}
    R^x(x_1,x_2:\mathcal{\beta},a_{\pm})
    :=
    \frac{1}{\mathcal{\beta}(x_1-x_2)^2}
    \Big(\frac{x_1x_2-\frac{a_-+a_+}{2}(x_1+x_2)+a_-a_+}{\sqrt{\sigma(x_1,a_{\pm})\sigma(x_2,a_{\pm})}}-1\Big),
\end{equation}
is the resolvent two-point function of the Dyson $\mathcal{\beta}$-ensemble, 
\begin{equation}
    \left(\Pi_idu_i\right)\left(\Pi_{1\leq i<j\leq L}|u_i-u_j|^{\beta}\right)e^{-L\frac{\beta}{2}\sum_iV(u_i)},
\end{equation}
with $\beta=2$. Importantly, $R^x(x_1,x_2:\mathcal{\beta},a_{\pm})$ depends on the potential $V$ only through the endpoints $a_{\pm}$ of the density of states. 

With (\ref{eq:relationtorandomstate}), we can obtain the continuous part of the fluctuation of the density of eigenvalues
\begin{eqnarray}\label{eq:density-two-point-non-tubular}
    \hat{D}(\lambda_1,\lambda_2)^{\text{Non-tubular}}
    &=&\frac{e^{2S(E)}}{2\pi^2(x_1-x_2)^2}\frac{x_1x_2-\frac{a_++a_-}{2}(x_1+x_2)+a_+a_-}{\sqrt{(x_1-a_-)(a_+-x_1)(x_2-a_-)(a_+-x_2)}}
    \nonumber\\
    &-&
    \frac{ke^{S(E)}}{4\pi^2}\frac{(x_1-\frac{a_++a_-}{2})(x_2-\frac{a_++a_-}{2})}{\sqrt{(x_1-a_-)(a_+-x_1)(x_2-a_-)(a_+-x_2)}}.
\end{eqnarray}
The first term comes from the fluctuation of unnormalized density matrices, and the second term comes from the fluctuation of the normalization. Thus we can write the fluctuations as 
\begin{eqnarray}\label{eq:nontubularexp}
    \hat{R}_{n_1n_2}^{\text{Non-Tubular}}
    =\int_{0}^1d\lambda_1d\lambda_2\lambda_1^{n_1}\lambda_2^{n_2}
    D(\lambda_1,\lambda_2)^{\text{Non-tubular}}.
\end{eqnarray}
The integral of 
$D(\lambda_1\lambda_2)_{\text{unnormalized}}$ in (\ref{eq:nontubularexp}) needs to be performed analytically in order to regulate a divergence coming from $1/(x_1-x_2)^2$.  

We can evaluate this integral in complete generality for certain values of $n$.  For example by explicitly expanding (\ref{eq:relationtorandomstate}), we find the Renyi entropy fluctuations
\begin{equation}
    (\delta \text{Tr}[\hat{\rho}_{\mathbf{R}}^{2}])^{\text{Non-Tubular}}
    =\frac{\sqrt{2}}{ke^{S(E)}},~
    (\delta\text{Tr}[\hat{\rho}_{\mathbf{R}}^{3}])^{\text{Non-Tubular}}
    =\frac{\sqrt{18k^2+39ke^{S(E)}+18e^{2S(E)}}}{k^2e^{2S(E)}}.
\end{equation}

We will now show that the nontubular contributions to the resolvent two-point function are identical to the resolvent two-point function of the random bipartite pure state of large dimensions $k$ and $e^{S(E)}$. Hence, the nontubular contributions to the entropy fluctuation are the same as the entropy fluctuation of the random biparatite pure state, which is known~\cite{Bianchi:2019stn}. 

%%%%%%%%%%%%%%%%%%%%%%%%%%%%%%%%%%%%%%%%%%%%%%%%%%%%%%%%%%%%%%%%%%%%%%%%%%%%%%%%%%%%%%%%%%%%%%%%%%%%%%%%%%%%%%%%%%%%%%%%%%%%%%%%%%%%%%%%%%%%%%%%%%%%%%%%%%%%%%%%%%%%%%%%%%%%%%%%%%%%%%%%

%\subsubsection{Resolvent of the Random Bi-partite Pure State}\label{subsection:matrixintegral}

Consider the Hilbert space $H=\mathbb{C}^m\otimes\mathbb{C}^n$ with $m\leq n$, and the random pure states in $H$ of the form
\begin{equation}\label{eq:randomstate}
    |\Psi\rangle=\sum_{1\leq i\leq m, 1\leq a\leq n}\Psi_{ia}|i\rangle\otimes|a\rangle.
\end{equation}
Here $\Psi_{ia}:=U_{(ia)(00)}$ is a matrix element with fixed column of a Haar-random unitary matrix $U\in U(mn)$. The matrix element of the reduced density matrix $\rho$ for the subsystem $\mathbb{C}^m$ is
\begin{equation}\label{eq:reduced}
    \langle i|\rho|j\rangle=\sum_a\Psi_{ia}\Psi_{ja}^*.
\end{equation}
%The important observation on the random states is that they are nearly maximally entangled on average

%%%%%%%%%%%%%%%%%%%%%%%%%%%%%%%%
%%%%%%%%%%%%%%%%%%%%%%%%%%%%%%%%%%%%%%%%%%%%
%%%%%%%%%%%%%%%%%%%%%%%%%%%%%%%%%%%%%%%%%%%%%%%%%%%
%%%%%%%%%%%%%%%%%%%%%%%%%%%%%%%%%%%%%%%%%%%%%%%%%%%%%%%%%%%%%%%%%%%%%%%%%%%%%%%%%%%%%%%%%%%%%%%%%%%%%%%%%%%%%%%%%%%%%%%%%%%%%%%%%%%%%%%%%%%%%%%%%%%%%%%%%%%%%%
%%%%%%%%%%%%%%%%%%%%%%%%%%%%%%%%%%%%%%%%%%%%%%%%%%%%%%%%%%%%%%%%%%%%%%%%%%%%%%%%%%%%%%%%%%%%%%%%%%%%%%%%%%%%%%%%%%%%%%%%%%%%%%%%%%%%%%%%%%%

Instead of treating matrix elements of Haar random unitary directly, we consider the Gaussian probability distribution for $\Psi_{ia}$
\begin{equation}\label{eq:prob}
    P(\Psi)\propto e^{-mn\Psi_{ia}\Psi_{ia}^*}.
\end{equation}
Note that the random state $|\Psi\rangle$ and the reduced density matrix $\rho$ are normalized only at their average, and thus their normalization can fluctuate. One might wonder if there are subleading contributions to (\ref{eq:prob}). This is not the case up to normalization for the following reason. Suppose we consider random unitary $U\in U(mn)$ and corresponding random complete orthonormal basis $|U_{(jb)}\rangle:=\sum_{1\leq i\leq m, 1\leq a\leq n}U_{(ia),(jb)}|i\rangle\otimes|a\rangle$. The standard way to obtain such random unitary matrix is to first consider $(nm)^2$ independent complex Gaussian random variables $\Psi_{ia}$ with zero mean and common variance (called Ginibre ensemble), and applying Gram-Schmidt orthonormalization to the vectors $\mathbf{v}_{(jb)}:=(\Psi_{(ia),(jb)})_{(ia)}$. The final set of vectors forms a complete orthonormal basis with unit probability, and the resulting measure on $U(mn)$ can be shown to be the Haar measure, see proposition 7.2 of \cite{Eaton} for proof. Since the first entry $|\Psi_{(11)}\rangle$ is the random pure state produced by the measure (\ref{eq:prob}) up to normalization, we can conclude that the entanglement spectrum of the random state can be simulated by (\ref{eq:prob}) after properly normalizing the state. We note that when we consider correlation between \emph{distinct} random state, the Gaussian approximation (\ref{eq:prob}) begins to deviate from the actual result. Nevertheless, it was shown in corollary 5 of \cite{Jiang2005} that for large $mn$, the first $mn/(\log mn)^2$ basis vectors can be treated as if they are independently chosen random states \footnote{More precisely, it was shown that the maximum of matrix element difference $mn|U_{(ia)(jb)}-\Psi_{(ia)(jb)}|$ among $mn/(\text{log} mn)^2$ entries goes to zero in the limit $mn\rightarrow\infty$.}.

From now on, we will use matrix integral to evaluate resolvent correlators.\footnote{The matrix integral portion this section was developed based on a note shared by Douglas Stanford, with his kind permission.} The ensemble induced by the probability distribution (\ref{eq:randomstate}) is known as complex Wishart ensemble or Laguerre $\beta=2$ ensemble, whose integration measure in terms of eigenvalues $\lambda_i$ of $\rho$ is given by \cite{Forrester+2010}
\begin{equation}
    e^{-mn\Psi_{ia}\Psi_{ia}^{*}}\Pi_{ia}d\Psi_{ia}d\Psi_{ia}^{*}
    \rightarrow
    C\left(\Pi_id\lambda_i\right)\left(\Pi_{1\leq i<j\leq m}(\lambda_i-\lambda_j)^2\right)\left(\Pi_i\lambda_i\right)^{n-m}e^{-mn\sum_i\lambda_i},
\end{equation}
where $C$ is a constant. After rescaling $x_i:=n\lambda_i$, we have an ordinary Altland-Zirnbauer matrix integral \cite{PhysRevB.55.1142} with $\beta=2$ and $\alpha=1$ (or the Dyson $\beta=2$), whose measure is
\begin{equation}
    \left(\Pi_idx_i\right)\left(\Pi_{1\leq i<j\leq m}|x_i-x_j|^{\beta}\right)\left(\Pi_i|x_i|^{\frac{\alpha-1}{2}}\right)e^{-m\frac{\beta}{2}\sum_iV(x_i)}
\end{equation}
with potential
\begin{equation}\label{eq:potential}
    V(x)=x-w\log x,~w=\frac{n-m}{m}.
\end{equation}
The leading order resolvent and the resolvent two-point correlator in $1/m$ expansion can be obtained via loop equations \cite{Stanford:2019vob}. For the resolvent $R^x(x)=R(\lambda=x/n)/(nm)$, the loop equation gives
\begin{equation}
    R^x(x)^2-V'(x)R^x(x)=\frac{-\frac{w}{m}\mathbb{E}[\sum_{i}\frac{1}{x_i}]}{x},
\end{equation}
at the leading in $m$. Combined with $R^x(x)\underset{x\rightarrow\infty}{\rightarrow}\frac{1}{x}$, we obtain 
\begin{equation}
    R(x)=nmR^x(x)=nm\left(\frac{1}{2}-\frac{w}{2x}-\frac{\sqrt{\frac{w^2}{4}-(1+\frac{w}{2})x+\frac{x^2}{4}}}{x}\right).
\end{equation}
In particular, $\mathbb{E}[\sum_ip_i]=1$.
The two-point correlator $R^x(x_1,x_2)$ of $R^x(x)$ is given by
\begin{equation}
    R^x(x_1,x_2)=n^2R^x(x_1,x_2:\beta=2,a_{\pm}=2+w\pm 2\sqrt{1+w}).
\end{equation}
So far, we have been considering the unnormalized reduced density matrix. We will now take into account the normalization and consider the normalized reduced density of state
\begin{equation}
    \langle i|\hat{\rho}|j\rangle=\frac{\sum_a\Psi_{ia}\Psi_{ja}^*}{\sum_{kb}\Psi_{kb}\Psi_{kb}^*}.
\end{equation}
To this end, we introduce a new coordinates, normalized eigenvalues $\hat{p_i}=p_i/r$ and $r=\sum_ip_i$. The integral measure is then
\begin{equation}
    dr\left(\Pi_id\hat{p}_i\right)\delta(1-\sum_i\hat{p}_i)r^{mn-1}
    \left(\Pi_{1\leq i<j\leq m}(\hat{p}_i-\hat{p}_j)^2\right)\left(\Pi_i\hat{p}_i\right)^{n-m}e^{-mnr}.
\end{equation}
Since $\mathbb{E}[\sum_ip_i]=1$, $r$ is peaked at $r=1$. The fluctuation is
\begin{equation}
    \mathbb{E}[(r-1)^2]=\frac{\int_0^{\infty} dr r^{mn-1}e^{-mnr}(r-1)^2}{\int_0^{\infty} dr r^{mn-1}e^{-mnr}}=\frac{1}{mn}.
\end{equation}
Thus we can treat the fluctuation of normalization perturbatively. At the leading order around $r=1$, we have
\begin{equation}
    \mathbb{E}[\text{Tr}[\hat{\rho}^{n_1}]\text{Tr}[\hat{\rho}^{n_2}]]-\mathbb{E}[\text{Tr}[\hat{\rho}^{n_1}]]\mathbb{E}[\text{Tr}[\hat{\rho}^{n_2}]]=
    \mathbb{E}[\text{Tr}[\rho^{n_1}]\text{Tr}[\rho^{n_2}]]
    -n_1n_2\mathbb{E}[\text{Tr}[\rho^{n_1}]]\mathbb{E}[\text{Tr}[\rho^{n_2}]]\mathbb{E}[(r-1)^2].
\end{equation}
Thus the connected resolvent two-point function defined in (\ref{eq:connectedtwopoint}) is given by, using (\ref{eq:twopoint}),
\begin{equation}
    \hat{R}(\lambda_1,\lambda_2)=R^x(x_1,x_2)-\frac{1}{mn}
    \partial_{x_1}(x_1R^x(x_1))
    \partial_{x_2}(x_2R^x(x_2))~.
\end{equation}
This is identical to (\ref{eq:relationtorandomstate}), the resolvent two-point function of the microcanonical PSSY model restricted to non-tubular wormholes, with $n=e^{S(E)}$ and $m=k$. %We thus can conclude that the random state model has the identical entanglement spectrum as the microcanonical PSSY model without tubular wormholes. 
Hence, the nontubular contributions to the entropy fluctuations in the PSSY model are the same as the entropy fluctuations in the random pure state~%
%\subsubsection{Entropy Fluctuation of the Random Pure State}
%The fluctuation in the entropy of the random pure state was studied in detail in 
\cite{Bianchi:2019stn}:
\footnote{The fluctuation is far smaller~\cite{Bianchi:2019stn} than a previously known upper bound~\cite{Patrick2006, Popescu:2006}:
\begin{equation}\label{eq:probability}
    \text{for}~3\leq m\leq n,~\text{Pr}\left(|S(\rho)-\mathbb{E}[S(\rho)]|\geq \delta\right)\leq 2\exp\left(-\frac{mn\delta^2}{36\pi^3(\log m)^2}\right),
\end{equation}
which leads to
\begin{equation}\label{eq:Hayden}    
    \delta S(\rho)\leq\frac{(6\pi)^{3/2}\log m}{\sqrt{mn}}.
\end{equation}
We note that this result is not a trivial consequence of the canonical typicality\cite{canonical,Popescu:2006}, because the combination of the Fannes inequality and $\mathbb{E}[|\rho_A-\mathbb{E}[\rho_A]|_1]\leq\sqrt{d_A/d_B}$ \cite{Popescu:2006} yields an estimate for the upper bound of the entropy fluctuation proportional to $\frac{d_A}{d_B}\log d_A$. We can also obtain an independent upper bound for the probability of the entropy fluctuation using a method given in the appendix of \cite{Cotler:2020jtk}. Indeed, since we have $-\log d_A\leq-\log\text{Tr}[\rho_A^2]\leq S_A$ in general, we can apply Markov's inequality to the non-negative variable $d_A\text{Tr}[\rho_A^2]-1$ to obtain $\text{Pr}[S\leq\log d_A-\delta]\leq \text{Pr}[e^{\delta}-1\leq d_A\text{Tr}\rho_A^2-1]\leq(d_A\langle\text{Tr}[\rho_A^2]\rangle-1)/(e^{\delta}-1)\approx(d_A/d_B)/(e^{\delta}-1)$, where we assumed large $d_A,d_B$ at the last approximate equality. Thus we obtain $\mathbb{E}[(S_A-\log d_A)^2]\lesssim 2\zeta(3)\,d_A^2/d_B^2$. This bound constrains the fluctuation around the average when $d_A\ll d_B$.}
\footnote{This result implies that the typical pure state $|\Psi\rangle$ on $\mathbb{C}^m\otimes \mathbb{C}^n:=H_A\otimes H_B$ fluctuates as
\begin{equation}\label{eq:ETHforentropy}
    S(\text{Tr}_{B}|\psi\rangle\langle \psi|)=\mathbb{E}[S_{A}]+\frac{1}{d_B}R_i,
\end{equation}
where $R_i$ is a real random variable with zero average with $O(1)$ variance. It would be interesting to relate this result (\ref{eq:ntu}) to the properties of energy eigenstates. The entropy and entropy variance for typical pure state and Gaussian pure state, which are models of eigenstates of chaotic and integrable system respectively, are given in \cite{Bianchi:2022}. It would also be interesting to relate this result to the eigenstate thermalization hypothesis (ETH) \cite{Deutsch:1991, Srednicki:1994, Rigol, singleigenstate, Dymarsky:,Lashkari:2016vgj}. The ETH states that any few-body operator $O$ and the majority of energy eigenstates $|E_i\rangle$ satisfy,  
\begin{equation}
    \langle E_i|O|E_j\rangle=O(E)^{\text{micro}}\delta_{ij}+e^{-S(\bar{E})/2}f(\bar{E},E_i-E_j)R_{ij},
\end{equation}
where $\bar{E}=(E_i+E_j)/2$, $f$ is a smooth real $O(1)$ function, and $R_{ij}$ is a random variable with unit variance and zero mean. The ETH for observables of a small subsystem, including entanglement entropy, was proposed in \cite{singleigenstate,Lashkari:2016vgj, Dymarsky:}. For few-body Hermitian operators it reduces to the random state prediction when we take a sufficiently small energy window of width given by the Thouless energy \cite{DAlessio:2015qtq}, by assuming that the transformation between eigenstates of the Hamiltonian and of $O$ is a random matrix. Yet it is not clear whether such an argument exists for entanglement entropy.}.

\begin{equation}\label{eq:ntu}
    \delta S(\rho)^{\text{Random Pure State}}
    =\left\{
    \begin{aligned}
    &\sqrt{\frac{1}{2n^2}-\frac{m}{4n^3}}+O(\frac{1}{nm^2})&  (m&< n)\\
    &\sqrt{\frac{1}{2m^2}-\frac{n}{4m^3}}+O(\frac{1}{mn^2})&  (m&> n),
    \end{aligned}
    \right.
\end{equation}
for large $m,~n$. Combined with the equivalence, this result implies that
\begin{eqnarray}\label{eq:ntu2}
    \delta S_{\mathbf{R}}^{\text{Non-tubular}}=
    \left\{
    \begin{aligned}
    &\sqrt{\frac{1}{2e^{2S(E)}}-\frac{k}{4e^{3S(E)}}}+O(\frac{1}{ke^{2S(E)}})&  (k&< e^{S(E)})\\
    &\sqrt{\frac{1}{2k^2}-\frac{e^{S(E)}}{4k^3}}+O(\frac{1}{k^2e^{S(E)}})&  (k&> e^{S(E)}),
    \end{aligned}
    \right.
\end{eqnarray}
We also obtain this result directly, albeit only for $k\gg e^{S(E)}$ and $k\ll e^{S(E)}$ in Appendix~\ref{appendix:perturbative}.

Another byproduct of the equivalence between the non-tubular PSSY model and the random pure state is the absence of any fluctuations of the rank of the density matrix. In the random pure state, this property is not restricted to the Gaussian approximation employed here. This is because in the space of pure states on $H=\mathbb{C}^m\otimes\mathbb{C}^n$, the subspace of states with rank strictly smaller than $m$ has lower dimensionality. Thus its measure is zero, and therefore the fluctuation vanishes. This is confirmed by explicit calculation~\cite{Bianchi:2019stn}. Thus we can conclude that 
%the fluctuation of the rank must be zero in the PSSY case without tubular wormhole at large $e^{S(E)}$, $k$ limit,
\begin{equation}\label{eq:rank}
    (\delta\text{Tr}[\hat{\rho}_{\mathbf{R}}^{0+}])^{\text{Non-Tubular}}
    =0~\text{for all}~k.
\end{equation}
However, in the next subsection we will see that tubular wormholes contribute a nonzero rank fluctuation in the PSSY model.

%%%%%%%%%%%%%%%%%%%%%%%%%%%%%%%%%%%%%%%%%%%%%%%%%%%%%%%%%%%%%%%%%%%%%%%%%%%%%%%%%%%%%%%%%%%%%%%%%%%%%%%%%%%%%%%%%%%%%%%%%%%%%%%%%%%%%%%%%%%%%%%%%%%%%%%%%%%%%%%%%%%%%%%%%%%%%%%%%%%%%%%%%%%%%%%%%%%%%%%%%%%%%%%%%%%%%%%%%%%%%%%%%%%%%%%%%%%%%%%%%%%%

%%%%%%%%%%%%%%%%%%%%%%%%%%%%%%%%%%%%%%%%%%%%%%%%%%%%%%%%%%%%%%%%%%%%%%%%%%%%%%%%%%%%%%%%%%%%%%%%%%%%%%%%%%%%%%%%%%%%%%%%%%%%%%%%%%%%%%%%%%%%%%%%%%%%%%%%%%%%%%%%%%%%%%%%%%%%%%%%%%%%%%%%%%%%%%%%%%%%%%%%%%%%%%%%%%%%%%%%%%%%%%%%%%

\subsection{Tubular Wormholes}\label{section:tubular}
\begin{figure}[t]
 \begin{center}
 \includegraphics[width=4cm,clip]{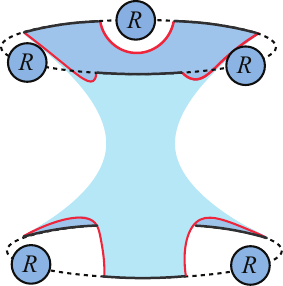}
 \end{center}
 \caption{Double trumpet connecting $R(\lambda_1)^{n}$ and $R(\lambda_2)^m$ in the first term of (\ref{eq:UUtubular}).}
 \label{fig:tubular}
 \end{figure}

%%%%%%%%%%%%%%%%%%%%%%%%%%%%%%%%%%%%%%%%%%%%%%%%%%%%%%%%%%%%%%%%%%%%%%%%%%%%%%%%%%%%%%%%%%%%%%%%%%%%%%%%%%%%%%%%%%%%%%%%%%%%%%%%%%%%%%%%%%%%%%%%%%%%%%%%%%%%%%%%%%%%%%%%%%%%%%%%%%%%%%%%%%%%
In this section, we consider contributions to $\delta S$ from tubular wormholes. Recall that tubular wormholes are defined as geometries that contain either double trumpets or handles. Since higher genus geometries are suppressed, we only need to consider geometries that contain a single double trumpet. Note that adding handles contributes to the fluctuation only at subleading order.
Using similar reasoning as the non-tubular case, we show in appendix \ref{appendix:proof} that 
\begin{eqnarray}
\label{eq:UUtubular}
    &&U(\lambda_1,\lambda_2)^{\text{Tubular}}
    \nonumber\\&=&
    \frac{1}{\lambda\bar{\lambda}}\sum_{n,m=1}^{\infty}\frac{Z^{(n,m)}_{\text{Double Trumpet}}R(\lambda_1)^{n}R(\lambda_2)^{m}}{nm(kZ_{\text{Disk}}^{(1)})^{n+m}}+
    R(\lambda_1)R(\lambda_2)
    \frac{k^2Z^{(1,1)}_{\text{Double Trumpet}}}{(kZ_{\text{Disk}}^{(1)})^2}
    \nonumber\\&-&
    \frac{1}{\lambda_1}\sum_{n_1=1}^{\infty}
    \frac{kZ^{(n,1)}_{\text{Double Trumpet}}R(\lambda_1)^{m}}{n(kZ_{\text{Disk}}^{(1)})^{n+1}}R(\lambda_2)
    -
    R(\lambda_1)
    \frac{1}{\lambda_2}\sum_{m=1}^{\infty}\frac{kZ^{(1,m)}_{\text{Double Trumpet}}R(\lambda_2)^{m}}{m(kZ_{\text{Disk}}^{(1)})^{n_2+1}}.\nonumber\\
\end{eqnarray}
The first term in \ref{eq:UUtubular} corresponds to the tubular wormhole exchanges between $\text{Tr}\rho_{\mathbf{R}}^{n_1}$ and $\text{Tr}\rho_{\mathbf{R}}^{n_2}$, see Fig \ref{fig:tubular}. 
The second and third terms correspond to those between $\text{Tr}\rho_{\mathbf{R}}^{n_1}$ and $(\text{Tr}[\rho_{\mathbf{R}}])^{n_2}$ etc. The fourth term to those between $(\text{Tr}[\rho_{\mathbf{R}}])^{n_1}$ and $(\text{Tr}[\rho_{\mathbf{R}}])^{n_2}$.

%%%%%%%%%%%%%%%%%%%%%%%%%%%%%
%%%%%%%%%%%%%%%%%%%%%%%%%%%%%%%%%%%%%%%%%%%%%%%%
%%%%%%%%%%%%%%%%%%%%%%%%%%%%%%%%%%%%%%%%%%%%%%%%%%%
%%%%%%%%%%%%%%%%%%%%%%%%%%%%%
%%%%%%%%%%%%%%%%%%%%%%%%%%%%%%%%%%%%%%%%%
%%%%%%%%%%%%%%%%%%%%%%%%%%%%%%%
%%%%%%%%%%%%%%%%%%%%%%%%%%%%
%%%%%%%%%%%%%%%%%%%%%%
%%%%%%%%%%%%%%%%%%%%%%%%%%%%%%%%
%%%%%%%%%%%%%%%%%%
%%%%%%%%%%%%%%%%%%%%%%
\paragraph{Microcanonical Ensemble}

For the microcanonical ensemble, the double trumpet partition function can be written as
\begin{equation}\label{eq:doubletrumpetmaintext}
    Z^{(n,m)}_{\text{Double Trumpet}}\approx\frac{\log (e^{\frac{3}{2}}\frac{\Delta E}{a})}{\pi^2}h(E,\mu)^{n+m},
\end{equation}
which was derived in appendix \ref{JTdt}. Using (\ref{eq:disk}) and (\ref{eq:doubletrumpetmaintext}), we can write explicitly
\begin{equation}
    \label{eq:microcononicalKP}
    U(\lambda_1,\lambda_2)^{\text{Tubular}}=
    \frac{\log (e^{\frac{3}{2}}\frac{\Delta E}{a})}{\pi^2\lambda_1\lambda_2}
    \left(\sum_{n=1}^{\infty}\frac{R(\lambda_1)^n}{n(ke^{S(E)})^n}-\frac{\lambda_1 R(\lambda_1)}{e^{S(E)}}\right)
    \left(\sum_{m=1}^{\infty}\frac{R(\lambda_2)^m}{m(ke^{S(E)})^m}-\frac{\lambda_2R(\lambda_2)}{e^{S(E)}}\right).
\end{equation}
The continuous part of the density two-point function is then
\begin{eqnarray}
    &&\hat{D}(\lambda_1,\lambda_2)_{\text{cont}}
    \nonumber\\&=&
    \frac{\log (e^{\frac{3}{2}}\frac{\Delta E}{a})}{\pi^2 e^{2S(E)}} 
    D(\lambda_1)_{\text{cont}}
    \left(1+\frac{x_1+w}{(w-x_1)^2-4x_1}\right)
    D(\lambda_2)_{\text{cont}}    \left(1+\frac{x_2+w}{(w-x_2)^2-4x_2}\right),\nonumber\\
\end{eqnarray}
where $D(\lambda)_{\text{cont}}$ is the continuous part of the density one point function. Writing $\alpha=k/e^{S(E)}$, we obtain
\begin{eqnarray}\label{eq:densitycor}
    \hat{R}_{n,m}^{\text{Tubular}}=
    \frac{\log (e^{\frac{3}{2}}\frac{\Delta E}{a})}{\pi^2 e^{2S(E)}}
    G(\alpha,n)G(\alpha,m)~,
\end{eqnarray}
where
\begin{eqnarray}\label{eq:Gfunction}
    &&G(\alpha,n)
    \nonumber\\&:=&
    \left(\frac{(1-\sqrt{\alpha})^2}{k}\right)^{n-1}
    \left({}_2F_1(-n+1,\frac{3}{2},3:-\frac{4\sqrt{\alpha}}{(1-\sqrt{\alpha})^2})
    \right.\nonumber\\
    &-&
    \left.\frac{1}{2}(1-\sqrt{\alpha})^2{}_2F_1(-n,\frac{1}{2},1:-\frac{4\sqrt{\alpha}}{(1-\sqrt{\alpha})^2})
    -\frac{1}{2}(1-\alpha){}_2F_1(-n+1,\frac{1}{2},1:
    -\frac{4\sqrt{\alpha}}{(1-\sqrt{\alpha})^2})\right).\nonumber\\
\end{eqnarray}
The expression (\ref{eq:Gfunction}) can have discontinuity at $\alpha=1$. Such discontinuity implies that the expression is no longer valid near $|k-e^{S(E)}|=O(1)$. For this reason, we will assume $|k-e^{S(E)}|\gg1$ in the following calculation.

Let us first consider the limiting cases $k\ll e^{S_0}$ and $k\gg e^{S_0}$. When $k\ll e^{S_0}$, the leading order terms are
\begin{eqnarray}\label{eq:smallktubular}
    &&\hat{R}_{n,m}^{\text{Tubular}}\rightarrow 
    \log (e^{\frac{3}{2}}\frac{\Delta E}{a})\frac{n(n-1)m(m-1)}{4\pi^2 e^{4S(E)}k^{n+m-4}},
    \nonumber\\&&
    (\delta S_{\mathbf{R}})^{\text{Tubular}}
    \rightarrow
    \frac{\sqrt{\log (e^{\frac{3}{2}}\frac{\Delta E}{a})}}{2\pi e^{2S(E)}k^{-1}},~
    (\delta \text{Tr}[\hat{\rho}_{\mathbf{R}}^{0+}])^{\text{Tubular}}
    \rightarrow 0.
\end{eqnarray}
When $k\gg e^{S(E)}$, the approximation (\ref{eq:largekapprox}) yields
\begin{eqnarray}\label{eq:largektubular}
    &&\hat{R}_{n,m}^{\text{Tubular}}
    \rightarrow
    \log (e^{\frac{3}{2}}\frac{\Delta E}{a})\frac{(n-1)(m-1)}{\pi^2 e^{(n+m)S(E)}},
    \nonumber\\&&
    (\delta S_{\mathbf{R}})^{\text{Tubular}}    
    \rightarrow
    \frac{\sqrt{\log (e^{\frac{3}{2}}\frac{\Delta E}{a})}}{\pi e^{S(E)}},~
    (\delta \text{Tr}[\hat{\rho}_{\mathbf{R}}^{0+}])^{\text{Tubular}}
    \rightarrow
    \frac{\sqrt{\log (e^{\frac{3}{2}}\frac{\Delta E}{a})}}{\pi}.
\end{eqnarray}
Next, we let $k,~e^{S(E)}$ be arbitrary but consider particular values of $n,~m$. Using the quadratic transformation of variables of hypergeometric functions and expanding around $n=m=1$, we obtain
\begin{equation}\label{eq:renyifluctuationtubular}
    \hat{R}_{n,m}^{\text{Tubular}}\underset{n,m\rightarrow 1}{\rightarrow}\left\{
    \begin{aligned}
    &\log (e^{\frac{3}{2}}\frac{\Delta E}{a})\frac{(n-1)(m-1)}{4\pi^2 e^{4S(E)}k^{n+m-4}}+O(k^{-(n+m+1)})&  &(1\ll e^{S(E)}-k)\\
    &\log (e^{\frac{3}{2}}\frac{\Delta E}{a})\frac{(1-\frac{e^{S(E)}}{2k})^2(n-1)(m-1)}{\pi^2 e^{(n+m)S(E)}}+O(e^
{-(n+m+1)S(E)})&  &(k-e^{S(E)}\gg1).
    \end{aligned}
    \right.
\end{equation}
We can confirm that (\ref{eq:renyifluctuationtubular}) reproduces (\ref{eq:smallktubular}) and (\ref{eq:largektubular}) for $n,~m\rightarrow 1$ in the limits. As a result, the entropy fluctuation is
\begin{equation}\label{eq:entropyfluctuationtubular}
    (\delta S_{\mathbf{R}})^{\text{Tubular}}=\left\{
    \begin{aligned}
    &\sqrt{\log (e^{\frac{3}{2}}\frac{\Delta E}{a})}\frac{k}{2\pi e^{2S(E)}}+O(e^{2S(E)}k^{-4})&  &(1\ll e^{S(E)}-k)\\
    &\sqrt{\log (e^{\frac{3}{2}}\frac{\Delta E}{a})}\frac{1-\frac{e^{S(E)}}{2k}}{\pi e^{S(E)}}+O(e^{-2S(E)})&  &(k-e^{S(E)}>1).
    \end{aligned}
    \right.
\end{equation}
A similar expansion around $n=m=0+$ yields
\begin{equation}\label{eq:rankfluctuationtubular}
    (\delta\text{Tr}[\hat{\rho}_{\mathbf{R}}^{0+}])^{\text{Tubular}}
    =\left\{
    \begin{aligned}
    &0+O\left(k^{-1}\right)&  &(1\ll e^{S(E)}-k)\\
    &\frac{\sqrt{\log (e^{\frac{3}{2}}\frac{\Delta E}{a})}}{\pi}+O\left(e^{-S(E)}\right)&  &(k-e^{S(E)}\gg1).
    \end{aligned}
    \right.
\end{equation}
For completeness, we note that the exact fluctuations for $n=2,~3$ are given by
\begin{equation}
    (\delta\text{Tr}[\hat{\rho}_{\mathbf{R}}^{2}])^{\text{Tubular}}
    =\frac{\sqrt{\log (e^{\frac{3}{2}}\frac{\Delta E}{a})}}{\pi e^{2S(E)}},~
    (\delta\text{Tr}[\hat{\rho}_{\mathbf{R}}^{3}])^{\text{Tubular}}
    =\sqrt{\log (e^{\frac{3}{2}}\frac{\Delta E}{a})}\frac{3+2\frac{k}{e^{S(E)}}}{\pi ke^{2S(E)}}.
\end{equation}

%%%%%%%%%%%%%%%%%%%%%%%%%%%%%%%%
%%%%%%%%%%%%%%%%%%%%%%%%%%%%%%%%%%%%%%%%%%%%
%%%%%%%%%%%%%%%%%%%%%%%%%%%%%%%%%%%%%%%%%%%%%%%%%%%
%%%%%%%%%%%%%%%%%%%%%%%%%%%%%%%%%
\subsection{Sum of Wormhole Contributions}\label{sum}

Summing up the two kinds of wormhole contributions, we obtain the total fluctuations of the entropy and of the rank of the Hawking radiation state. We assume large $k$ and $e^{S(E)}$ with $ke^{-S(E)}$ fixed, as well as $|k-e^{S(E)}|\gg1$. Then
\begin{eqnarray}\label{eq:summary1}
    \delta S_{\mathbf{R}}=e^{-S(E)}\times\left\{
    \begin{aligned}
    &\sqrt{\frac{1}{2}-\frac{k}{4e^{S(E)}}+\frac{\log (e^{\frac{3}{2}}\frac{\Delta E}{a})}{4\pi^2}\frac{k^2}{e^{2S(E)}}}+O(e^{2S(E)}k^{-3})&  &(1\ll e^{S(E)}-k)\\
    &\sqrt{\frac{e^{2S(E)}}{2k^2}-\frac{e^{3S(E)}}{4k^3}+\frac{\log (e^{\frac{3}{2}}\frac{\Delta E}{a})}{\pi^2}\left(1-\frac{e^{S(E)}}{2k}\right)^2}+O(e^{-S(E)})&  &(k-e^{S(E)}\gg1).
    \end{aligned}
    \right.
\end{eqnarray}
\begin{equation}\label{eq:summary2}
    \delta\text{Tr}[\hat{\rho}_{\mathbf{R}}^{0+}]
    =\left\{
    \begin{aligned}
    &~\,0 ~~ +O\left(k^{-1}\right)&  &(1\ll e^{S(E)}-k)\\
    & \frac{1}{\pi}\sqrt{\log (e^{\frac{3}{2}}\frac{\Delta E}{a})}+O\left(e^{-S(E)}\right)&  &(k-e^{S(E)}\gg1)~.
    \end{aligned}
    \right.
\end{equation}
The behavior of the rank fluctuation can be explained in the following manner. When $k<e^{S(E)}$, the rank of the total pure state (\ref{eq:state}) is $k$ which is maximal and thus is insensitive to the fluctuation of the bulk Hilbert space dimension $e^{S(E)}$. On the other hand, when $k>e^{S(E)}$, the rank is $e^{S(E)}$ and receives corrections directly from the fluctuation of $e^{S(E)}$. 
%%%%%%%%%%%%%%%%%%%%%%%%%%%%%%%%
%%%%%%%%%%%%%%%%%%%%%%%%%%%%%%%%%%%%%%%%%%%%
%%%%%%%%%%%%%%%%%%%%%%%%%%%%%%%%%%%%%%%%%%%%%%%%%%%
%%%%%%%%%%%%%%%%%%%%%%%%%%%%%%%%%%%%%%%%%%%%%%%%%%%%%%%%%%%%%%%%%%%%%%%%%%%%%%%%%%%%%%%%%%%%%%%%%%%%%%%%%%%%%%%%%%%%%%%%%%%%%%%%%%%%%%%%%%%%%%%%%%%%%%%%%%%%%%
%%%%%%%%%%%%%%%%%%%%%%%%%%%%%%%%%%%%%%%%%%%%%%%%%%%%%%%%%%%%%%%%%%%%%%%%%%%%%%%%%%%%%%%%%%%%%%%%%%%%%%%%%%%%%%%%%%%%%%%%%%%%%%%%%%%%%%%%%%%

\section{Discussion}\label{section:random state}

The entropy fluctuation (\ref{eq:summary1}) is exponentially suppressed in the black hole entropy at all times. At first this is surprising: like the Page curve itself, the fluctuations should be symmetric under exchange of the Hilbert space dimension of the two subsystems. 

What breaks this symmetry is the fact that the black hole is prepared in the microcanonical ensemble, whose Hilbert space dimension can vary between different theories in the ensemble of theories. By contrast, the radiation Hilbert space has fixed dimension $k$. 

Once the black hole becomes the smaller subsystem --- after the Page time --- its Hilbert space size controls both the Page curve and its fluctuations. The entropy is approximately given by the log of the rank in this regime. The nonzero rank fluctuation (\ref{eq:summary2}) in the microcanonical ensemble dominates over the much smaller entropy fluctuation expected if both Hilbert space dimensions were fixed. To see this, substitute (\ref{eq:summary2}) into (\ref{eq:Pagecurve}); this yields (\ref{eq:summary1}).

A fluctuation of the bulk Hilbert space dimension could be modeled by a generalization of the random state model. Consider the direct product of factorized Hilbert spaces $H=(\oplus_{a}\mathbb{C}^{m_a})\otimes\mathbb{C}^n$, in the state
\begin{equation}  |\Psi\rangle=\sum_{a}\sqrt{p_a}|\Psi_a\rangle,~\text{where}~|\Psi_a\rangle\in\mathbb{C}^{m_a}\otimes\mathbb{C}^n,~\sum_ap_a=1.
\end{equation}
Here $|\Psi_a\rangle$ is a random state in $\mathbb{C}^{m_a}\otimes\mathbb{C}^{n}$, and $p_a$ obeys an appropriate probability distribution. The reduced density matrix on $\mathbb{C}^{n}$, for $m_a<n$, is the probabilistic sum of the reduced density matrices for $|\Psi_a\rangle$, whose rank is $m_a$. The average entropy is $\log m_a-\frac{m_a}{2n}$. If $p_a$ is sufficiently peaked as a function of $m_a$ at some value $m_a=\bar{m}$, this generalized model will reproduce the fluctuations of the rank and the entropy observed in the PSSY model.

%%%%%%%%%%%%%%%%%%%%%%%%%%%%%%%%
%%%%%%%%%%%%%%%%%%%%%%%%%%%%%%%%%%%%%%%%%%%%
%%%%%%%%%%%%%%%%%%%%%%%%%%%%%%%%%%%%%%%%%%%%%%%%%%%
%%%%%%%%%%%%%%%%%%%%%%%%%%%%%%%%%%%%%%%%%%%%%%%%%%%%%%%%%%%%%%%%%%%%%%%%%%%%%%%%%%%%%%%%%%%%%%%%%%%%%%%%%%%%%%%%%%%%%%%%%%%%%%%%%%%%%%%%%%%%%%%%%%%%%%%%%%%%%%
%%%%%%%%%%%%%%%%%%%%%%%%%%%%%%%%%%%%%%%%%%%%%%%%%%%%%%%%%%%%%%%%%%%%%%%%%%%%%%%%%%%%%%%%%%%%%%%%%%%%%%%%%%%%%%%%%%%%%%%%%%%%%%%%%%%%%%%%%%%

%%%%%%%%%%%%%%%%%%%%%%%%%%%%%%%%%%%%%%%%%%%%%%%%%%%%%%%%%%%%

\paragraph{Acknowledgements}
We are grateful to Douglas Stanford for initial collaboration, and for generously sharing results on the nontubular contributions to the fluctuations in the PSSY model. We thank Eugenio Bianchi for making us aware of Ref.~\cite{Bianchi:2019stn}, where the fluctuation of the entropy of the random pure state was analytically computed. We also thank L.~Iliesiu, Y.~Nakata, G.~Penington, and T.~Takayanagi for discussions and correspondence. RB was supported in part by the Berkeley Center for Theoretical Physics; by the Department of Energy, Office of Science, Office of High Energy Physics under QuantISED Award DE-SC0019380 and under contract DE-AC02-05CH11231; and by the National Science Foundation under Award Number 2112880.

%%%%%%%%%%%%%%%%%%%%%%%%%%%%%%%%%%%%%%%%%%%%%%%%%%%%%%%%%%%%%%%%%%%%%%%%%%%%%%%%%%%%%%%%%%%%%%%%%%%%%%%%%%%%%%%%%%%%%%%%%%%%%%%%%%%%%%%%%%%%%%%%%%%%%%%%%%%%%%%%%%%%%%%%%%%%%%%%%%%%%%%%%%%%%

%%%%%%%%
%%%
%%%%%%%%%%%%%%%%%%%%%%%%
\appendix
\section{Renyi Entropy Fluctuations}
\label{appfluc}
In this appendix, we explain the details of the computation of Renyi entropy fluctuations.
%%%%%%%%%%%%%%%%%%%%%%%%%%%%%%%%%%%%%%%%%%%%%%%%
%%%%%%%%%%%%%%%%%%%%%%%%%%%%%%%%%%%%%%%%%%%%%%%%%%
%%%%%%%%%%%%%%%%%%%%%%%%%%%%%%%%%%%%%%%%%%%%
%%%%%%%%%%%%%%%%%%%%%%%%
\subsection{Non-tubular Wormhole}
\label{appflucnt}
 
In this appendix, we compute contributions to $\hat{R}_{n_1n_2}$ from contributions of second, third and forth terms in Eq.~(\ref{eq:UUnontubular}), which we write as $\hat{R}^{\text{Non-tubular}}_{n_1n_2~\text{Normalization}}$. We can perform the integration directly in this case and find explicit contribution to the entropy fluctuation for any value of $ke^{-S(E)}$. 
We have,
\begin{eqnarray}
    \hat{R}^{\text{Non-tubular}}_{n_1n_2~\text{Normalization}}:=\int_{0}^1d\lambda_1d\lambda_2\lambda_1^{n_1}\lambda_2^{n_2}
    D(\lambda_1,\lambda_2)^{\text{Non-tubular}}_{\text{Normalization}},
\end{eqnarray}
where
\begin{equation}
    \hat{D}(\lambda_1,\lambda_2)^{\text{Non-tubular}}_{\text{Normalization}}:=-
    \frac{ke^{S(E)}}{4\pi^2}\frac{(x_1-\frac{a_++a_-}{2})(x_2-\frac{a_++a_-}{2})}{\sqrt{(x_1-a_-)(a_+-x_1)(x_2-a_-)(a_+-x_2)}},
\end{equation}
which is the second term of Eq.~(\ref{eq:density-two-point-non-tubular}).

We can explicitly see that
\begin{equation}
    \hat{R}^{\text{Non-tubular}}_{n_1n_2~\text{Normalization}}=
    \frac{-e^{S(E)}}{4k}F(\alpha,n_1)F(\alpha,n_2),
\end{equation}
where $\alpha=k/e^{S(E)}$ and
\begin{eqnarray}
    F(\alpha,n)=
    \left\{
    \begin{aligned}
    &\frac{1}{k^n}\left({}_2F_1(-n-1,-n-1,1,\alpha)
    -
    (1+\alpha){}_2F_1(-n,-n,1,\alpha)\right) 
    &  (\alpha<1)&\\
    &\frac{1}{e^{nS(E)}}
    \left(\alpha{}_2F_1(-n-1,-n-1,1,\frac{1}{\alpha})
    -
    (1+\alpha){}_2F_1(-n,-n,1,\frac{1}{\alpha})\right) &(\alpha>1)&,
    \end{aligned}
    \right.
\end{eqnarray}
Near $n_1,n_2\rightarrow 1$, we  have
\begin{eqnarray}
    \hat{R}^{\text{Non-tubular}}_{nm\text{Normalization}}\rightarrow
    \left\{
    \begin{aligned}
    &-\frac{\left(1+(\frac{\alpha}{2}+1)(n-1)\right)
    \left(1+(\frac{\alpha}{2}+1)(m-1)\right)}{k^{n+m-1}e^{S(E)}}
    &  (\alpha<1)&\\
    & -\frac{\left(1+(\frac{1}{2\alpha}+1)(n-1)\right)
    \left(1+(\frac{1}{2\alpha}+1)(m-1)\right)}{ke^{(n+m-1)S(E)}}
    &(\alpha>1)&.
    \end{aligned}
    \right.\nonumber\\
\end{eqnarray}
And for $n_1=n_2=0+$, 
\begin{eqnarray}
    \hat{R}^{\text{Non-tubular}}_{0+0+~\text{Normalization}}=0.
\end{eqnarray}

%%%%%%%%%%%%%%%%%%%%%%%%%%%%%%%%%%%%%%%%%%%%%%
%%%%%%%%%%%%%%%%%%%%%%%%%%%%%%%%%%%%%%%%%%%%%%%%%%
%%%%%%%%%%%%%%%%%%%%%%%%%%%%%%%%%%%%%%%%%%%%
%%%%%%%%%%%%%%%%%%%%%%%%
\subsection{Tubular Wormhole}
\label{appfluct}

This subsection completes the derivation of eq. (\ref{eq:renyifluctuationtubular}) from (\ref{eq:Gfunction}), by studying the behavior of the function (\ref{eq:Gfunction}) at $n\rightarrow 1$. We first invoke the quadratic transformation of variables of hypergeometric function 
\begin{equation}
    {}_2F_1(a,b,a-b+1:z)=(1-\sqrt{z})^{-2a}{}_2F_1(a,a-b+\frac{1}{2},2a-2b+1:-\frac{4\sqrt{z}}{(1-\sqrt{z})^2}),~|z|<1.
\end{equation}
We can apply this relation to $G(\alpha,n)$ for $z=\alpha$ assuming $\alpha<1$, obtaining 
\begin{eqnarray}
    G(\alpha,n)
    =
    \frac{2{}_2F_1(1-n,-n,2:\alpha)-{}_2F_1(-n,-n,1:\alpha)- 
    (1-\alpha){}_2F_1(1-n,1-n,1,\alpha)}{2k^{n-1}}.\nonumber\\
\end{eqnarray}
Applying for $z=1/\alpha$ assuming $\alpha>1$ yields
\begin{eqnarray}
     &&G(\alpha,n)\nonumber\\&&
     =\frac{2{}_2F_1(1-n,-n,2:1/\alpha)-\alpha{}_2F_1(-n,-n,1:1/\alpha)- 
    (1-\alpha){}_2F_1(1-n,1-n,1,1/\alpha)}{2e^{(n-1)S(E)}}.\nonumber\\
\end{eqnarray}
Using these identities we can compute the fluctuation $R_{n_1n_2,\text{Tubular}}$ for general replica number. For expression near $n_1=n_2=1$, we use expansions such as ${}_2F_1(-n+1,-n,2:\alpha)=1+\frac{\alpha}{2} (n-1)+O((n-1)^2)$, ${}_2F_1(-n,-n,1:\alpha)=1+\alpha+2\alpha (n-1)+O((n-1)^2)$ and ${}_2F_1(-n+1,-n+1,1:\alpha)=1+O((n-1)^2)$. Using these relations, we arrive at (\ref{eq:renyifluctuationtubular}).

%%%%%%%%%%%%%%%%%%%%%%%%%%%%%%%%%%
%%%%%%%%%%%%%%%%%%%%%%%%%%%%%%%%%%%%%%%%%%%%%%%%%%%
%%%%%%%%%%%%%%%%%%%%%%%%%%%%%%%%%%%%%%%%%%%%%%%%%%
%
%%%%%%%%%%%%%%%%%%%%%%%%%%%%%%%%%%%%%%%%%%%%%%%%%%%%%%%%%%%%%%%%%%%%
%%%%%%%%%%%%%%%%%%%%%%%%%%%%%%%%%%%%%%%%%%%%%%%%%%%%%%%%%%%%%%%%%%%%%%%%%%%%%%%%%%%%%%%%%
%%%%%%%%%%%%%%%%%%%%%%%%%%%%%%%%%%%%%%%%%%%%%%%%%%%%%%%%%%%%%%%%%%%%%%%
%%%%%%%%%%%%%%%%%%%%%%%%%%%%%%%%%%%%%%%%%%%%%%%%%%%%%%%%%%%%%%%%%%%%%%%%%%%%%%%%%%%%%%%%%%

\section{JT Gravity Partition Functions}\label{appendix:JT}

In this appendix, we explain some results in JT gravity \cite{Teitelboim:1983ux, Jackiw:1984je, Maldacena:2016upp, Stanford:2017thb, Yang:2018gdb,Harlow:2018tqv,Saad:2019lba,Stanford:2019vob, Saad:2019pqd} used in this paper. The main objectives are the microcanonical partition functions (\ref{eq:disk}) and (\ref{eq:doubletrumpetmaintext}).

\subsection{Disk}\label{JTdisk}

This subsection completes the derivation of eq. (\ref{eq:disk}). We first consider the disk partition function. The normalized density of states of the disk is
\begin{equation}
    D_{\text{Disk}}(E)=\frac{\sinh(2\pi\sqrt{2E})}{2\pi^2}.
\end{equation}
The disk partition function is
\begin{equation}
    Z_{\text{Disk}}(\beta)
    =
    e^{S_0}\int_0^{\infty}dE D_{\text{Disk}}(E)e^{-\beta E}
    =
    e^{S_0}\frac{e^{\frac{2\pi^2}{\beta}}}{\sqrt{2\pi}\beta^{\frac{3}{2}}}.
\end{equation}
The Hartle-Hawking wave function for a disk with a boundary segment $x$ and geodesic boundary with length $l$ is \cite{Yang:2018gdb, Saad:2019pqd}
\begin{equation}
    \psi_{\text{Disk}}(x,l)=\int_0^{\infty}dED_{\text{Disk}}(E)\psi_{\text{Disk}}(E,l)e^{-xE},
\end{equation}
where
\begin{equation}
    \psi_{\text{Disk}}(E,l)=4e^{-l/2}K_{i\sqrt{8E}}(4e^{-l/2}).
\end{equation}
The Hartle-Hawking wave function satisfies
\begin{equation}
    \int_{-\infty}^{\infty}\frac{dle^{l}}{2}~\psi_{\text{Disk}}(\beta_1,l)\psi_{\text{Disk}}(\beta_2,l)
    =
    \int_0^{\infty}dED_{\text{Disk}}(E)e^{-(\beta_1+\beta_2)E}
    =
    Z_{\text{Disk}}(\beta_1+\beta_2),
\end{equation}
where we used
\begin{equation}
    \int_{-\infty}^{\infty}dl~K_{i\sqrt{8E}}(4e^{-l/2})K_{i\sqrt{8E'}}(4e^{-l/2})=\frac{\delta(E-E')}{8D_{\text{Disk}}(E)}.
\end{equation}
The path integral of a disk with $n$ geodesic boundaries with lengths $l_1,\cdots,l_n$ is
\begin{equation}
    I_{n}(l_1,\cdots,l_n)=2^{n}\int_0^{\infty}dED_{\text{Disk}}(E)K_{i\sqrt{8E}}(4e^{-l_1/2})\cdots K_{i\sqrt{8E}}(4e^{-l_n/2}).
\end{equation}
This satisfies
\begin{eqnarray}
    &&    
    e^{l/2}\psi_{\text{Disk}}(x_1+\cdots+x_{n-1},l)
    \nonumber\\&=&
    2\int_{-\infty}^{\infty}dl_1...dl_{n-1}~e^{\frac{l_1+...+l_{n-1}}{2}}I_{n}(l_1,...,l_{n-1},l)\psi_{\text{Disk}}(x_1,l_1)...\psi_{\text{Disk}}(x_{n-1},l_{n-1}),
\end{eqnarray}
and
\begin{eqnarray}
    &&    
    Z_{\text{Disk}}(x_1+\cdots+x_n)
    =
    \int_{-\infty}^{\infty}dl_1...dl_{n}~e^{\frac{l_1+...+l_{n}}{2}}I_{n}(l_1,...,l_{n})\psi_{\text{Disk}}(x_1,l_1)...\psi_{\text{Disk}}(x_{n},l_{n}).\nonumber\\
\end{eqnarray}

The path integral for a disk with $n$ geodesic boundaries with lengths $l_1,\cdots,l_n$ and $n$ conformal boundary segments with lengths $x_1,...,x_n$ is
\begin{eqnarray}
    &&e^{(l_1+\cdots+l_n)/2}\psi_{\text{Disk}}(x_1,...,x_n:l_1,...,l_n)
    \nonumber\\&=&
    2^n\int_{-\infty}^{\infty}dl_1'...dl_n'~e^{(l_1'+\cdots+l_n')/2}I_{2n}(l_1,...,l_1',....)\psi_{\text{Disk}}(l_1',x_1)...\psi_{\text{Disk}}(l_n',x_n)
    \nonumber\\&=&
    2^{2n}\int_0^{\infty}dED_{\text{Disk}}(E)K_{i\sqrt{8E}}(4e^{-l_1/2})... K_{i\sqrt{8E}}(4e^{-l_n/2})e^{-(x_1+....+x_n)E}.\nonumber\\
\end{eqnarray}
Replacing the $n$ geodesic boundaries by $n$ EOW branes with action $S_{\text{EOW}_i}=\mu l_i$, we obtain the bulk partition function
\begin{eqnarray}
    &&Z^{(n)}_{\text{Disk}}\left[\text{canonical}, \text{boundary lengths}=x_i\right]
    \nonumber\\
    &=&\int_{-\infty}^{\infty}dl_1...dl_n~e^{l_1+\cdots+l_n}
    \psi_{\text{Disk}}(x_1,...,x_n:l_1,...,l_n)e^{-\mu(l_1+\cdots+l_n)}    
    \nonumber\\
    &=&e^{S_0}\int_0^{\infty}dED_{\text{Disk}}(E)h(E,\mu)^ne^{-(x_1+....+x_n)E},\nonumber\\
\end{eqnarray}
where
\begin{eqnarray}
    h(E,\mu):=2^{2}\int_{-\infty}^{\infty} dle^{l/2}K_{i\sqrt{8E}}(4e^{-l/2})e^{-\mu l}
    =\frac{|\Gamma(\mu-1/2+i\sqrt{2E})|^2}{2^{2\mu-1}},
\end{eqnarray}
for $\text{Re}[\mu]-\frac{1}{2}-|\text{Im}[\sqrt{8E}]|>0$.

%%%%%%%%%%%%%%%%%%%%%%%%%%%%%%%%%%%%%%%%%%%%%%%%%%%%%%%%%%%%%%%%%%%%%%%%%%%%%%%%%%%%%%%%%%%%%%%%%%%%%%%%%%%%%%%%%%%%%%%%%%%%%%%%%%%%%
\subsection{Double Trumpet}\label{JTdt}

This subsection completes the derivation of eq. (\ref{eq:doubletrumpetmaintext}). A tubular wormhole can be divided into two trumpets along a geodesic. We first consider a single trumpet bounded by a geodesic with length $b$. The path integral for fixed geodesic boundary $b$ and AdS boundary segment with length $x$, and a geodesic boundary anchored from the AdS boundary with  length $l$ is given by \cite{Saad:2019pqd}
\begin{equation}
    \psi_{\text{Trumpet}}(l,x,b)=\int_0^{\infty}dED_{\text{Trumpet}}(E,b)\psi_{\text{Disk}}(E,l)e^{-xE},
\end{equation}
where
\begin{equation}
    D_{\text{Trumpet}}(E,b)=\frac{\text{cos}(b\sqrt{2E})}{\pi\sqrt{2E}}.
\end{equation}
We first consider the partition function without EOW branes. We define 
\begin{eqnarray}\label{eq:Ddoubletrumpet}
    D_{\text{Double Trumpet}}(E,E')&:=&
    \underset{d\rightarrow 2}{\lim}\int_{0}^{\infty}b^{d-1}db~D_{\text{Trumpet}}(E,b)D_{\text{Trumpet}}(E',b)
    \nonumber\\&=&
    -\frac{E+E'}{4\pi^2\sqrt{EE'}(E-E')^2}.
\end{eqnarray}
Note that it follows that 
\begin{equation}
     \int_0^{\infty}dED_{\text{Double Trumpet}}(E,E')=\int_0^{\infty}dE'D_{\text{Double Trumpet}}(E,E')=0.
\end{equation}
Defining
\begin{equation}
    D_{\text{Double Trumpet}}(E,E':d):=-\frac{E+E'}{4\pi^2\sqrt{EE'}(E-E')^d},
\end{equation}
the partition function of the double trumpet with boundary length $x$ and $y$ is then given by
\begin{eqnarray}
    Z_{\text{Double Trumpet}}(x:y)&=&\underset{d\rightarrow 2}{\lim}\int_0^{\infty}dEdE'D_{\text{Double Trumpet}}(E,E':d)e^{-xE-yE'}=\frac{\sqrt{xy}}{2\pi(x+y)}.\nonumber\\
\end{eqnarray}

Next we generalize to the partition function with an arbitrary number of AdS boundaries and geodesic boundaries anchored from them. The path integral for a trumpet Hartle-Hawking wavefunction with $n$ geodesic boundaries with length $l_i~(i=1,\cdots, n)$ and AdS boundary segments of length $x_a~(a=1,...,n)$ can be written as
\begin{eqnarray}
    &&e^{(l_1+\cdots+l_n)/2}\psi_{\text{Trumpet}}(l_1,...,l_n:x_1,...,x_n:b)\nonumber\\
    &=&
    2^n\int_{-\infty}^{\infty}
    dl_1'...dl_n'~e^{(l'_1+....+l'_n)/2}I_{2n}(l_1,...,l_n,l_1',...,l_n')\psi_{\text{Trumpet}}(l_1',x_1,b)
    \psi_{\text{Disk}}(l_2',x_2)...\psi_{\text{Disk}}(l_n',x_n)
    \nonumber\\&=&
    2^{2n}\int_0^{\infty}dED_{\text{Trumpet}}(E,b)K_{i\sqrt{8E}}(4e^{-l_1/2})\cdots K_{i\sqrt{8E}}(4e^{-l_n/2})e^{-(x_1+...+x_n)E}.
\end{eqnarray}
When the geodesics of this path integral are replaced by EOW branes, we have
\begin{eqnarray}
    Z_{\text{Trumpet}}^{(n)}(x_1,...,x_n:b)
    &=&
    \int_{-\infty}^{\infty}dl_1...dl_n~e^{l_1+\cdots+l_n}
    \psi_{\text{Trumpet}}(x_1,...,x_n:l_1,...,l_n:b)e^{-\mu(l_1+\cdots+l_n)}
    \nonumber\\
    &=&
    \int_0^{\infty}dED_{\text{Trumpet}}(E,b)h(E,\mu)^ne^{-(x_1+...+x_n)E}.
\end{eqnarray}

Next, we consider a single tubular wormhole exchange between two disk topology $n$- and $m$-boundary partition functions. It is given by
\begin{eqnarray}
    &&Z^{(n,m)}_{\text{Double Trumpet}}(x_1,...,x_n:y_1,...,y_m)
    \nonumber\\&=&
    \int_{0}^{\infty}bdb~Z_{\text{Trumpet}}(x_1,...,x_n:b)Z_{\text{Trumpet}}(y_1,...,y_m:b)\nonumber\\&=&\int_0^{\infty}dEdE'D_{\text{Double Trumpet}}(E,E')e^{-(x_1+...)E-(y_1+...)E'}
    h(E,\mu)^nh(E',\mu)^m.
\end{eqnarray}

The microcanonical double trumpet $(n,m)$-boundary partition function is
\begin{eqnarray}
    &&
    Z^{(n,m)}_{\text{Double Trumpet}}\left[\text{microcanonical}, \text{energy}=E,~E', \text{width}=\Delta E\right]
    \nonumber\\
    &:=&
    \Pi_{i=1}^{n+m}
    \left[
    \int_0^{\infty}dE_i f_{(E,\Delta E,a)}(E_i)
    \int_{x_i\in \gamma+i\mathbb{R}}dx_i
    e^{x_iE_i}\right]
    Z^{(n,m)}_{\text{Double Trumpet}}(x_1,...,x_n:x_{n+1},...,x_{n+m})
    \nonumber\\&=&\int_0^{\infty}d\tilde{E}d\tilde{E}'
    D_{\text{Ramp}}(\tilde{E},\tilde{E}')
    \left(f_{(E,\Delta E,a)}(\tilde{E})\right)^n
    \left(f_{(E',\Delta E,a)}(\tilde{E}')\right)^m
    h(\tilde{E},\mu)^nh(\tilde{E}',\mu)^m.
\end{eqnarray}
Here we assume $a\ll \Delta E$ but $1\ll e^{S(E)}\frac{a}{\Delta E}=e^{S_0}D_{\text{Disk}}(E)a$, thus $a$ is of order $1$ quantity in $e^{S_0}$. We introduced the continuous smearing function for microcanonical ensemble. The sharp top hat cannot be used in the approximation we use in this paper, where higher genus geometries are neglected. The continuous smearing function we use in this paper is
\begin{equation}
    f_{(E,\Delta E,a)}(\tilde{E})
    =\left\{
    \begin{aligned}
    &-\frac{\tilde{E}-(E+\Delta E/2+a)}{a}& &(E+\Delta E/2<\tilde{E}<E+\Delta E/2+a)\\
    &1&  &(E-\Delta E/2<\tilde{E}<E+\Delta E/2 )\\
    &\frac{\tilde{E}-(E-\Delta E/2-a)}{a}&  &(E+\Delta E/2-a<\tilde{E}<E-\Delta E)\\
    &0&  &(\text{otherwise}).
    \end{aligned}
    \right.
\end{equation}
In the work of PSSY \cite{Penington:ReplicaWormholeWestCoast}, such regularization was unnecessary; thus the sharp window could be used.

Let us compute the microcanonical double trumpet $(n,m)$-boundary partition function. By utilizing an approximation 
\begin{eqnarray}
    D_{\text{Double Trumpet}}(E,E')\approx
    \frac{-1}{2\pi^2(E-E')^2},
\end{eqnarray}
we can perform the integration explicitly, arriving at
\begin{eqnarray}\label{eq:doubletrumpet}
    \label{eq:ramp1}Z^{(n,m)}_{\text{Double Trumpet}}
    \left[\text{microcanonical}, \text{energy}=E, \text{width}=\Delta E\right]    \approx\frac{\log (e^{\frac{3}{2}}\frac{\Delta E}{a})}{\pi^2}h(E,\mu)^{n+m}.\nonumber\\
\end{eqnarray}

So far we have studied the partition function for a trapezoid smearing function. The purpose was to regularize the divergence of (\ref{eq:Ddoubletrumpet}), which is an artifact of cutting off higher genus contributions in the GPI. In the following, for reference, we assume that the density of state two-point function is given by the sine-kernel
\begin{equation}
    D_{\text{Sine-Kernel}}(E,E')=
    -\frac{1}{\pi^2}\left(\frac{\sin^2(e^{S_0}(E-E'))}{(E-E')^2}-\pi e^{S_0}\delta(E-E')\right),
\end{equation}
which reduces to (\ref{eq:Ddoubletrumpet}) after smearing with width larger than $e^{-S_0}$. The partition function in this case is finite. Assuming that $\Delta E$ is an $O(1)$ quantity, so that $e^{S_0}\Delta E$ is large, we then obtain
\begin{eqnarray}
    \label{eq:ramp2}Z^{(n,m)}_{\text{Sign Kernel}}
    \left[\text{microcanonical}, \text{energy}=E, \text{width}=\Delta E\right]    \approx\frac{\log (2e^{1+\gamma}e^{S_0}\Delta E)}{\pi^2}h(E,\mu)^{n+m}.\nonumber\\
\end{eqnarray}
This result is similar to (\ref{eq:doubletrumpet}) when we take $a\approx e^{-S_0}$, consistent with the expectation that higher genus contributions in the GPI regularize the spectrum at the scale $e^{-S_0}$.

%%%%%%%%%%%%%%%%%%%%%%%%%%%%%%%%%%%%%%%%%%%%%%%%%%%%%%%%%%%%%%%%%%%%%%%%%%%%%%%%%%%%%%%%%%%%%%%%%%%%%%%%%%%%%%%%%%%%%%%%%%%%%%%%%%%%%%%%%%%%%%%%%%%%%%%%%%%%%%%%%%%%%%%%%%%%%%%%%%%%%%%%%%%%%%%%%%%%%%%%%%%%%%%%%%%%%%%%%%%%%%%%%%%%%%%%%%%%%%%%%%%%%%%%%%%%%%%%%%%%%%%%%%%%%%%%%%%%%%%%%%%%%%%%%%%%%%%%%%%%%%%%%%%%%%%%%%%%%%%%%%%%%%%%%%%%%%%%%%%%%%%%%%%%%%%%%%%%%%%%%%%%%%%%%%%%%%%%%%%%%%%%%%%%%%%%%%%%%%%%%%%%%%%%%%%%%%%%%%%%%%%%%%%%%%%%%%%%%%%%%%%%%%%%%%%%%%%%%%%%%%%%%%%%%%%%%%%%%%%%%%%%%%%%%%%%%%%%%%%%%%%%%%%%%%%%%%%%%%%%%%%%%%%%
\section{Proofs of (\ref{eq:UUnontubular}) and (\ref{eq:UUtubular})}\label{appendix:proof}

In this appendix, we provide the proofs for (\ref{eq:UUnontubular}) and (\ref{eq:UUtubular}).

\subsection{Proof of (\ref{eq:UUnontubular})}

Since the derivation of the second to fourth terms of (\ref{eq:UUnontubular}) is straightforward, we will only show that the multiplicity of the first term in (\ref{eq:UUnontubular}) correctly captures the number of geometries. First, we can classify any diagram by the number $n$ of irreducible ladders. There are two possibilities: that there is no cyclic symmetry in the diagram, and that there is.

Let us first consider the case when there is no cyclic symmetry in the diagram. We denote the number of boundaries of these diagrams by $n_1$ and $n_2$. In the first term of (\ref{eq:UUnontubular}), such a diagram appears $n$ times; therefore, the factor $\frac{1}{n}$ in (\ref{eq:UUnontubular}) cancels this multiplicity. Since this diagram appears in $R(\lambda_1,\lambda_2)$ $n_1n_2$ times, we conclude that the counting for diagrams without cyclic symmetry is correctly captured.

Next, we consider the case when there is a cyclic symmetry with periodicity $(s_1, s_2)$. Then the diagram can be decomposed into $m=n_1/s_1=n_2/s_2$ identical copies. Each copy contains $n/m$ irreducible ladders. Therefore, in (\ref{eq:UUnontubular}), the diagram appears $n/m$ times. By the factor $1/n$ in (\ref{eq:UUnontubular}), the coefficient in (\ref{eq:UUnontubular}) for the diagram is now $\frac{1}{m}$. On the other hand, the diagram appears in $ R(\lambda_1,\lambda_2)$ 
\begin{equation}
    m\times s_1 \times s_2=\frac{1}{m}\times n_1n_2 
\end{equation}
times, matching precisely with the counting of (\ref{eq:UUnontubular}). Thus the counting of diagrams with cyclic symmetry is also correctly captured in (\ref{eq:UUnontubular}).

\subsection{Proof of (\ref{eq:UUtubular})}

The multiplicity of the first term in (\ref{eq:UUtubular}) can be checked in a similar manner as (\ref{eq:UUnontubular}). First, we can classify diagrams by cyclic symmetry of the diagram for each side $\text{Tr}\rho^{n_1}$ and $\text{Tr}\rho^{n_2}$.
Let us suppose the periodicity of $\text{Tr}\rho^{n_1}$ and $\text{Tr}\rho^{n_2}$ are $(s_1, s_2)$. Here we include $s_1=n_1$ or $s_2=m_2$ for the case where there is no cyclic symmetry. $(n,m)$ corresponds to the number of boundaries that are connected to the double trumpet. Thus in the first term of (\ref{eq:UUtubular}), the diagram appears $ns_1/n_1\times ms_2/n_2$ times. By the factor $1/(nm)$, the coefficient for the diagram in (\ref{eq:UUtubular}) is $s_1s_2/(n_1n_2)$. On the other hand, the diagram appears in $R(\lambda_1,\lambda_2)$ $s_1s_2$ times, matching precisely with the counting of (\ref{eq:UUtubular}). Thus we conclude that the counting of diagrams with cyclic symmetry is also correctly captured in (\ref{eq:UUtubular}).

%%%%%%%%%%%%%%%%%%%%%%%%%%%%%%%%%%%%%%%%%%%%%%%%%%%%%%%%%%%%%%%%%%%%%%%%%%%%%%%%%%%%%%%%%%%%%%%%%%%%%%%%%%%%%%%%%%%%%%%%%%%%%%%
\section{Perturbative Computation}\label{appendix:perturbative}

In this appendix, we evaluate the entropy fluctuation $\delta S_{\mathbf{R}}$ for non-tubular wormholes for small and large values of $k/e^{S(E)}$. A simpler but related integral was studied in \cite{Forrester:2012} near $k\sim e^{S(E)}$ (see also \cite{Forrester:2004}). It would be interesting to compare them with the explicit evaluation in \cite{Bianchi:2019stn}. 

When $k\ll e^{S_0}$, we can use the approximation 
\begin{equation}\label{eq:smallkapprox}
    \lambda R(\lambda)-k\rightarrow\frac{1}{\lambda}+\frac{1}{k\lambda^2}+\frac{1}{k^2\lambda^3}+\frac{1}{k^3\lambda^4}+\cdots=\sum_{n=1}^{\infty}\frac{1}{k^{n-1}\lambda^{n}}.
\end{equation}
Substituting this into (\ref{eq:relationtorandomstate}) yields 
\begin{equation}
    \hat{R}_{n_1n_2}^{\text{Non-Tubular}}
    \rightarrow
    \frac{n_1(n_1-1)n_2(n_2-1)}{2k^{n_1+n_2-2}e^{2S(E)}}.
\end{equation}
The fluctuation of the rank and the entropy are
\begin{equation}\label{eq:smallk}
    (\delta S_{\mathbf{R}})^{\text{Non-Tubular}}\rightarrow\frac{1}{\sqrt{2}e^{S(E)}},~(\delta\text{Tr}[\hat{\rho}_{\mathbf{R}}^{0+}])_{\text{Non-Tubular}}\rightarrow 0.
\end{equation}

For $k\gg e^{S(E)}$, we can approximate
\begin{equation}\label{eq:largekapprox}
    \lambda R(\lambda)-k\rightarrow\frac{1}{\lambda}+\frac{1}{e^{S(E)}\lambda^2}+\frac{1}{e^{2S(E)}\lambda^3}+\frac{1}{e^{3S(E)}\lambda^4}+\cdots=\sum_{n=1}^{\infty}\frac{1}{e^{(n-1)S(E)}\lambda^{n}},
\end{equation}
which leads to 
\begin{equation}
    \hat{R}_{n_1n_2}^{\text{Non-Tubular}}
    \rightarrow
    \frac{n_1(n_1-1)n_2(n_2-1)}{2e^{(n_1+n_2-2)S(E)}k^2}.
\end{equation}
The fluctuation of the rank and the entropy are
\begin{equation}\label{eq:largek}
    (\delta S_{\mathbf{R}})^{\text{Non-Tubular}}\rightarrow\frac{1}{\sqrt{2}k},~(\delta\text{Tr}[\hat{\rho}_{\mathbf{R}}^{0+}])^{\text{Non-Tubular}}\rightarrow 0.
\end{equation}
To summarize, we obtain at large $k$, $e^{S(E)}$
\begin{equation}
    (\delta S_{\mathbf{R}})^{\text{Non-Tubular}}=\left\{
    \begin{aligned}
    &\frac{1}{\sqrt{2}k}&  (k&\ll e^{S(E)})\\
    &\frac{1}{\sqrt{2}e^{S(E)}}&  (k&\gg e^{S(E)}).
    \end{aligned}
    \right.
\end{equation}

%%%%%%%%%%%%%%%%%%%%%%%%%%%%%%%%%%%%%%%%%%%%%%%%%%%%%%%%%%%%%%%%%%%%%%%%%%%%%%%%%%%%%%%%%%%%%%%%%%%%%%%%%%%%%%%%%%%%%%%%%%%%%%%%%%%%%%%%%

\bibliographystyle{JHEP}
\bibliography{Main.bib}
\end{document}